\documentclass[showpacs,superscriptaddress,groupedaddress] {revtex4-1}         

\usepackage{graphicx}                           
\usepackage{amsmath}
\usepackage{amssymb}
\usepackage{dsfont}
\usepackage{dcolumn}    
\usepackage{color}                                                      
\usepackage{bm}
\usepackage{braket}

\usepackage{marginnote}

\usepackage{booktabs}
\usepackage{array}
\newcolumntype{C}{>{$}c<{$}}

\DeclareMathAlphabet{\mathpzc}{OT1}{pzc}{m}{it}

\newcommand{\ud}{\mathrm{d}}
\newcommand{\Z}{\mathds{Z}}
\newcommand{\ui}{\mathrm{i}}

\newcommand{\ISSUE}[1]{\textcolor{red}{\textbf{[#1]}}}
\newcommand{\RESOLVED}[1]{\textcolor{blue}{\textbf{[#1]}}}

\newcommand{\rem}[1]{{}}

\begin{document}

\title{Quantum Monodromy in the Isotropic 3-Dimensional Harmonic Oscillator}

\author{Irina Chiscop$^1$, Holger R. Dullin$^2$, Konstantinos Efstathiou$^1$, and Holger Waalkens$^1$\\
\small $^1$Bernoulli Institute for Mathematics, Computer Science and Artificial Intelligence, University of Groningen\\$^2$School of Mathematics and Statistics, University of Sydney}

\date{\today}

\begin{abstract}
The isotropic harmonic oscillator in dimension 3 separates in several different coordinate systems.
Separating in a particular coordinate system defines a system of three commuting operators,
one of which is the Hamiltonian.
We show that the joint spectrum of the Hamilton operator, 
the $z$ component of the angular momentum, and a quartic integral 
obtained from separation in prolate spheroidal coordinates 
has quantum monodromy for sufficiently large energies.
This means that one cannot  globally assign quantum numbers to the joint spectrum.
The effect can be classically explained by showing that the corresponding 
Liouville integrable system has a non-degenerate focus-focus point, and hence Hamiltonian monodromy.
\end{abstract}

\maketitle


\section{Introduction}

The isotropic harmonic oscillator is at the same time the simplest and the most important system in physics.
The system is very special in both the classical and the quantum setting.  All (nontrivial) solutions of the classical equations of motion are periodic and even have the same period. The quantum system is special in that it has an equidistant energy spectrum. 
The best explanation of these special properties in both the classical and the quantum setting 
are the symmetries of the system. 
The energy spectrum is independent of the dimension, however, the degeneracy of the energy levels increases with dimension.
What we are going to show is that within a degenerate energy eigenspace 
we can define a \emph{quantum integrable system} (QIS)  whose joint spectrum is non-trivial in the sense that it does not allow for a global assignment of  quantum numbers.
%
With a QIS for an $N-$dimensional isotropic harmonic oscillator we mean a set of 
 $N$ commuting operators $\mathcal{H} = (\hat H_1, \dots, \hat H_N)$ with say  $\hat H_1$ being the Hamilton operator of the system. 
Because the operators commute their spectra can be measured simultaneously:
$\hat H_i \psi  = \lambda_{i} \psi$, $i=1, \dots, N$.
Together they define the  joint spectrum which associates a point in $N-$dimensional space with coordinates $\lambda_{i}$ to each eigenfunction $\psi$.
It follows from the Bohr-Sommerfeld quantization of classical actions whose local existence in turn follows  from the Liouville-Arnold Theorem \cite{Arnold78}
that the joint spectrum locally has the structure of a lattice $\Z^N$. 
We show that for $N=3$ there is a QIS for which there is an obstruction to the global existence of action-angle variables due to monodromy \cite{Duistermaat80},
which manifests itself as a lattice defect in the joint spectrum that prevents the global assignment of quantum numbers \cite{CushDuist88,VuNgoc99,SadovskiiZhilinskii99,Zhilinskii2006}.
Monodromy and generalizations of monodromy \cite{Nekhoroshev2006,Sadovskii2007164} have been extensively studied in recent years and found for many different systems, see, e.g., \cite{Zhilinskii2011} and the references therein. Quantum monodromy explains, e.g., problems in assigning rovibrational spectra of molecules \cite{CDGHJLSZ04,Child2008,Assematetal2010} 
or electronic spectra of atoms in external fields \cite{CushmanSadovskii99,EfstathiouSadovskiiZhilinskii2007}.
Moreover it provides a mechanism for excited-state quantum phase transitions \cite{Cejnaretal2006,Caprio20081106}.
The generalization of monodromy to scattering systems has been shown to lead to defects in the lattice of transparent states
in planar central scattering \cite{DullinWaalkens08}. 
Monodromy can also play a role in spatiotemporal nonlinear wave systems \cite{Sugnyetal2009}, and
dynamical manifestations of monodromy have recently been studied in \cite{Delos14}.


Another way of thinking about our result is as follows. Due to the high degree of symmetry the 
quantum harmonic oscillator is not only a QIS but it has additional independent operators 
that commute with $\hat H$. Such a system is called super-integrable.
%
%
Important examples  are systems that are separable in different coordinate systems.
Schwarzschild \cite{Schwarzschild1916} %
was the first to point out that if the Hamilton-Jacobi equation of $H$ can be separated in more 
than one coordinate system, the quantum energy eigenvalues of $\hat H$ are degenerate.
Such a Hamiltonian operator $\hat H$ is called multiseparable, 
and is hence included in non-equivalent QIS's $\mathcal H$ and $\mathcal G$. 
The simplest multiseparable systems are the free particle, the Kepler problem, and  the harmonic oscillator. 
A multiseparable system with $N$ degrees of freedom is superintegrable, because if both 
$\mathcal{H}$ and $\mathcal{G}$ contain $\hat H$, then we have found 
more than $N-1$ operators that commute with $\hat H$.
An  important group  of 3-dimensional superintegrable and multiseparable systems
is classified in \cite{Kress06}.

The classical geometry of superintegrable systems is well understood.  Fixing the integrals defines tori of lower dimension than in the Liouville-Arnold Theorem and Nekhoroshev showed that one can construct lower dimensional action-angle coordinates in a kind of generalization of the Liouville-Arnold Theorem \cite{Nekhoroshev1972}. 
More global aspects have been studied in \cite{FomenkoMishchenko78,Delzant87}.
The isotropic three-dimensional harmonic oscillator is maximally superintegrable which means that together with  the Hamiltonian it has five independent integrals. The joint level sets are one-dimensional tori, i.e. periodic orbits, whose projection to configuration space are ellipses centered at the center of the force. 
From the classical geometric point of view considering tori with half the dimension of phase space in 
a super-integrable system appears somewhat arbitrary. However, from the quantum point of view it 
is prudent to study all possible sets of commuting observables, because these tell us what can 
be measured simultaneously as the uncertainty principle is trivial in this case.
Thus we are going to study a particular set of collections of ellipse shaped periodic orbits that form 3-tori in phase space, 
and we will show that the joint quantum spectrum associated to these tori has quantum monodromy.


If a Hamiltonian $\hat H$ is super-integrable then there are distinct 
QIS that share the given Hamiltonian $\hat H$, but form non-equivalent 
QIS with in general different joint spectra. The eigenvalues of $\hat H$ and their degeneracy 
are the same in each realisation, but the joint spectrum within a degenerate eigenspace
and the corresponding basis of eigenfunctions are different. 

We are focusing on the case where the different QIS are obtained from 
separation in different coordinate systems. 
Separation in different coordinate systems gives different QIS with the same Hamiltonian $\hat H$. 
A Hamiltonian that is multi-separable is also super-integrable, since there are more than $n$ integrals. 
For the 3-dimensional harmonic oscillator this is well known.
On the one hand it separates in Cartesian coordinates into a sum of one-degree-of-freedom 
harmonic oscillators,  so that the wave function for the multi-dimensional case is simply a product
of wave functions for the one-dimensional case, which are given in terms of Hermite polynomials.
On the other hand it separates in spherical coordinates, which leads to wave functions that 
are products of spherical harmonics and associated Laguerre polynomials. 
The associated quantum numbers have different meaning, but the total number
of states of a three-dimensional harmonic oscillator with angular frequency $\omega$ and energy $E =\hbar \omega( n + 3/2)$ is $(n+1)(n+2)/2$ with ``principalÕÕ qauntum number $n=0,1,2,\ldots$.
In the first case we have a quantum number $n_i = 0, 1, 2,\dots$ for each 1D oscillator,
and the eigenvalues of $\hat H$ are $E = \hbar \omega (n + 3/2) =  \hbar \omega (n_1 + n_2 + n_3 + 3/2)$.
In the second case (see, e.g., \cite{Griffiths2016}) 
we have $E = \hbar \omega (2k+l+3/2)$ for non-negative integer $k$ where $l$ is the total angular momentum 
eigenvalue $l = n, n-2, n-4, \dots$ down to 0 or 1, depending on whether $n$ is even or odd, respectively.
In addition there is the usual ``magneticÕÕ quantum number $m = -l, \dots, l$.
In both cases the quantum states form a lattice in which lattice points can be uniquely labelled 
by quantum numbers. The details of the two lattices are, however, different. In particular the actions are not even locally related by unimodular transformation.   

Specifically, we are going to separate the isotropic harmonic oscillator in prolate spheroidal coordinates.
Prolate spheroidal coordinates are a family of coordinate systems where the family parameter $a$
is half the distance between the focus points of a family of confocal ellipses and hyperbolas, which in order to get corresponding coordinate surfaces are rotated about 
the axis containing the  focus points. 
In the limit $a \to 0$ spherical coordinates are obtained, and in the limit $a \to \infty$ parabolic 
coordinates are obtained.
Our main result is that when the energy $E > \frac12 \omega^2 a^2$ then the system has monodromy. 
Our approach is similar to a recent analysis of the Kepler problem \cite{DullinWaalkens2018},
which through separation in prolate spheroidal coordinates leads to a quantum integrable system
that does not possess three global quantum numbers. 

This paper is organized as follows. In Sec.~\ref{sec:separation} we introduce the classical three-dimensional isotropic harmonic oscillator, discuss its symmetries and its separation in prolate spheriodal coordinates.
In Sec.~\ref{sec:bifdiag} we compute the bifurcation diagram for the energy momentum map associated with separation in prolate spheroidal coordinates and prove the presence of monodromy.
The effect of monodromy on the  quantum spectrum is studied in Sec.~\ref{sec:sepQM}.
We conclude with some comments in Sec.~\ref{sec:discussion}.


\section{Classical separation in prolate spheroidal coordinates} 
\label{sec:separation}


The three-dimensional isotropic harmonic oscillator  has Hamiltonian
\begin{equation}\label{eq:defH}
    H = \frac12 |\mathbf{p}|^2 + \frac{\omega^2}{2}  |\mathbf{r}|^2 \,,
\end{equation}
where $\mathbf{r} = (x,y,z)^T$ 
and $\mathbf{p} = (p_x, p_y, p_z)^T$  are the canonical variables on the phase space $T^*\mathbb{R}^3\cong \mathbb{R}^6$.
By choosing suitable units we can assume that the frequency $\omega$ has the value $1$.
But in order to identify terms arising from the potential we will keep $\omega$ in the equations below.
Not only are the three separated Hamiltonians 
$$\mathbf{A} = ( \tfrac12 (p_x^2 + \omega^2 x^2), \tfrac12 (p_y^2 + \omega^2 y^2), \tfrac12 (p_z^2 + \omega^2 z^2) )^T $$ 
constants of motion, but so are the components of the angular momentum 
$\mathbf{L} = \mathbf{r} \times \mathbf{p}$.
Not all these integrals are independent. But any 
five of them are, so that $H$ is maximally superintegrable.

Define $$\mathbf{B} = ( \{L_x, A_y\} ,  \{L_y, A_z\} ,  \{L_z, A_x\}  )^T,$$
where $\{ \cdot \,,\,\cdot \}$ is the Poisson bracket.  The algebra of 9 quadratic integrals 
$\mathbf{A, B, L}$ closes and defines a Lie-Poisson bracket, shown in Table~\ref{eq:poisson-cp2}, that is isomorphic to the Lie algebra $\mathfrak{su}(3)$ (see also \cite{Fradkin1965}). Fixing the relations between the integrals $\mathbf{A}$, $\mathbf{B}$, $\mathbf{L}$ defines an embedding of the reduced symplectic manifold $\mathbb{C}P^2$ into $\mathbb{R}^9$.
Here $\mathbb{C}P^2$ is the orbit space of the $S^1$ action induced on $\mathbb C^3 \simeq T^* \mathbb R^3$ by the Hamiltonian flow of $H$ \cite{Moser1970}. 
The Hamiltonian $H = A_x + A_y + A_z $ is a Casimir. The algebra has two more Casimirs, the quadratic $C_2 = 2 \mathbf{A}^2 + \omega^2 \mathbf{L}^2 + \mathbf{B}^2$ and the cubic
\[
C_3 = 6 \mathrm{Re}(w_x w_y w_z) + \sum_{k=x,y,z} 2|w_k|^2(H-3A_k) - \frac{8}{27}(H-3A_k)^3,
\]
where $w_k = B_k + i \omega L_k$, $k=x,y,z$.


\begin{table}[htbp]
\renewcommand{\arraystretch}{1.2}
\begin{ruledtabular}
\begin{tabular}{C|CCCCCCCCC}
\{\downarrow,\rightarrow\} & A_x & A_y & A_z & L_x & L_y & L_z & B_x & B_y & B_z \\
\colrule
 A_x & 0 & 0 & 0 & 0 & B_y & -B_z & 0 & -\omega ^2 L_y & \omega ^2 L_z \\
 A_y & 0 & 0 & 0 & -B_x & 0 & B_z & \omega ^2 L_x & 0 & -\omega ^2 L_z \\
 A_z & 0 & 0 & 0 & B_x & -B_y & 0 & -\omega ^2 L_x & \omega ^2 L_y & 0 \\
 L_x & 0 & B_x & -B_x & 0 & L_z & -L_y & 2 A_z-2 A_y & -B_z & B_y \\
 L_y & -B_y & 0 & B_y & -L_z & 0 & L_x & B_z & 2 A_x-2 A_z & -B_x \\
 L_z & B_z & -B_z & 0 & L_y & -L_x & 0 & -B_y & B_x & 2 A_y-2 A_x \\
 B_x & 0 & -\omega ^2 L_x & \omega ^2 L_x & 2 A_y-2 A_z & -B_z & B_y & 0 & -\omega ^2 L_z & \omega ^2 L_y \\
 B_y & \omega ^2 L_y & 0 & -\omega ^2 L_y & B_z & 2 A_z-2 A_x & -B_x & \omega ^2 L_z & 0 & -\omega ^2 L_x \\
 B_z & -\omega ^2 L_z & \omega ^2 L_z & 0 & -B_y & B_x & 2 A_x-2 A_y & -\omega ^2 L_y & \omega ^2 L_x & 0 \\
\end{tabular}
\end{ruledtabular}
\caption{Poisson structure on $\mathbb{C}P^2$.\label{eq:poisson-cp2}}
\end{table}

The huge symmetry of the isotropic harmonic oscillator is also reflected by its separability in different coordinate systems. 
In fact, the three-dimensional oscillator separates in several different coordinate systems. The most well known are the systems of Cartesian coordinates and spherical coordinates (see, e.g., \cite{Fradkin1965}). 
In this paper we will be studying the separation in prolate spheroidal coordinates. The separability in these coordinates is, e.g., mentioned in \cite{CoulsonJoseph1967}. The coordinates are defined  with respect to 
two focus points which we assume to be located on the $z$ axis at $\mathbf{a}=(0,0,a)$ and $-\mathbf{a}=(0,0,-a)$ where $a>0$.    
The prolate spheroidal coordinates are then defined as
$$
(\xi,\eta,\varphi) = \Big(\frac{1}{2a} ( r_+ + r_-),\, \frac{1}{2a} (r_+ - r_-),\, \mathrm{arg}(x+\ui y) \Big) ,
$$
where $r_\pm = |\mathbf{r}\pm \mathbf{a}|$. They have ranges $\xi\ge1$, $-1\le \eta\le1$ and $0\le \varphi \le 2\pi$. 
The surfaces of constant $\xi>1$ and $-1<\eta<1$ are confocal prolate ellipsoids and two-sheeted hyperboloids which are rotationally symmetric about the $z$ axis and have focus points at $\pm \mathbf{a}$. 
For $\xi\to 1$, the ellipsoids collapse to the line segment connecting the focus points, 
and for $\eta\to \pm 1$, the hyperboloids collapse to the half-lines consisting of the part of the $z$ axis above and below the focus points, respectively.
 
The Hamiltonian in prolate spheroidal coordinates becomes
$$
H = \frac12 \frac{1}{a^2(\xi^2-\eta^2)} (p_\xi^2(\xi^2-1) + p_\eta^2 (1-\eta^2)) + \frac{1}{2} \frac{p_\varphi^2}{a^2(\xi^2-1)(1-\eta^2)}  + \frac12 a^2 \omega^2 (\xi^2 + \eta^2-1).
$$
The angle $\varphi$ is cyclic. So $p_\varphi$ which is the $z$ component of the angular momentum is a constant of motion. 
Multiplying the energy equation $H=E$ by $2a^2(\xi^2-\eta^2)$ and reordering terms gives the separation constant
\begin{equation} \label{eq:G_xieta}  
\begin{split}
G &:= - p_\xi^2(\xi^2-1) -  \frac{l_z^2}{\xi^2-1}   - a^4 \omega^2 \xi^2 (\xi^2 -1) + 2a^2 (\xi^2-1)E\\
&\phantom{:}=
\phantom{-} p_\eta^2 (1-\eta^2) +  \frac{l_z^2}{1-\eta^2} + a^4 \omega^2 \eta^2(1-\eta^2) - 2a^2 (1-\eta^2)E,
\end{split}
\end{equation}
where we use $l_z$ to denote the value of $p_\varphi$. 
Rewriting the separation constant in Cartesian coordinates gives

\begin{equation}\label{eq:G_Cart} 
G =  L_x^2 + L_y^2 + L_z^2  - 2 a^2 (A_x + A_y).
\end{equation}
The functions  $\mathcal{G} = ( H, L_z, G)$  are independent and their mutual Poisson brackets vanish. They thus define a Liouville integrable system which as we will see has a singular foliation by Lagrangian tori with monodromy which we then also study quantum mechanically. 


\section{Bifurcation diagram and reduction}
\label{sec:bifdiag}

Solving \eqref{eq:G_xieta} for the momenta $p_\eta$ and $p_\xi$ we get
\begin{equation} \label{eq:separated_momenta}
p^2_\xi = \frac{P(\xi)}{(\xi^2-1)^2} \quad \text{ and }  \quad p^2_\eta = \frac{P(\eta)}{(\eta^2-1)^2} \,,
\end{equation}
where 
\begin{equation} \label{eq:def_P(s)} 
P(s) = -l_z^2 + 2 a^2 (1-s^2) \left[ \left(E - \frac12 a^2 \omega^2 s^2 \right) (1-s^2) + \frac{g}{2a^2} \right]
\end{equation}
with $g$ denoting the value of the separation constant $G$.
The roots of the polynomial $P(s)$ are turning points in the corresponding separated degree of freedom, i.e. roots in $[-1,1]$ correspond to turning points in the $(\eta , p_\eta)$ phase plane and roots  in $[1,\infty)$ correspond to turning points in the $(\xi, p_\xi)$ phase plane. Critical motion occurs for values of the constants of motion where turning points collide, i.e. for double-roots of $P(s)$. 
The bifurcation diagram, i.e. the set of critical values of the energy momentum map $\mathcal{G} = ( H, L_z, G): T^*\mathbb{R}^3 \to \mathbb{R}^3$, $(\mathbf{r} ,\mathbf{p} )\mapsto (E,l_z,g)$, can thus  be found from 
the vanishing of the discriminant of the polynomial $P(s)$. However, care has to be taken due to the singularities of the prolate spheroidal coordinates at the focus points. 
In Sec.~\ref{sec:reduction} below we will therefore derive the bifurcation diagram more rigorously using the method of singular reduction \cite{CushBates}. 
For $l_z=0$, the motion (in configuration space) takes place in invariant planes of constant angles about the $z$ axis. 
We will consider this case first and study the case of 
general $l_z$ afterwards.


\subsection{The two-dimensional harmonic oscillator ($l_z = 0$)}

\begin{figure}
\includegraphics[width=4.5cm]{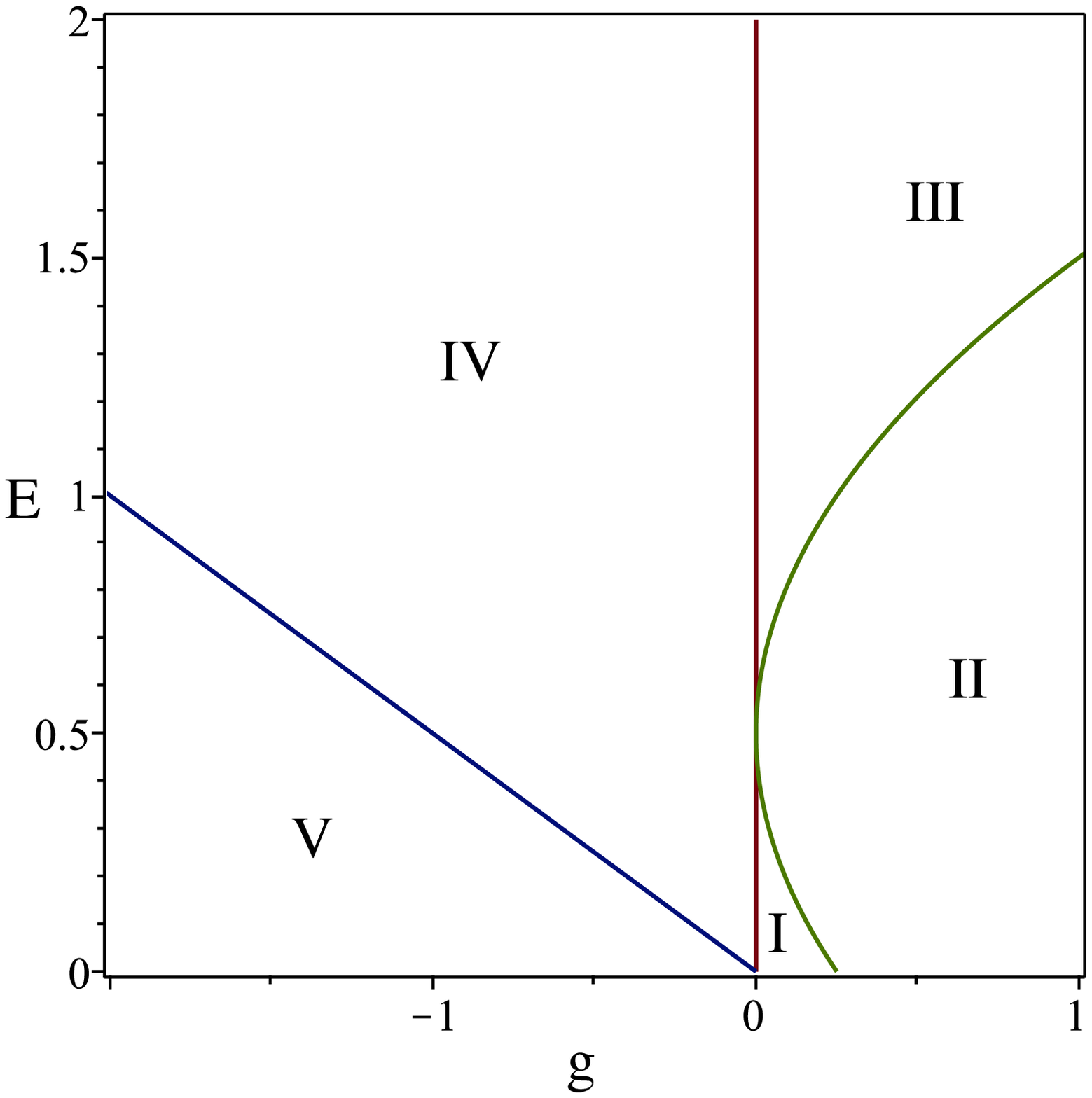} \hspace*{1cm}
\raisebox{3cm}{I}
\includegraphics[width=4cm]{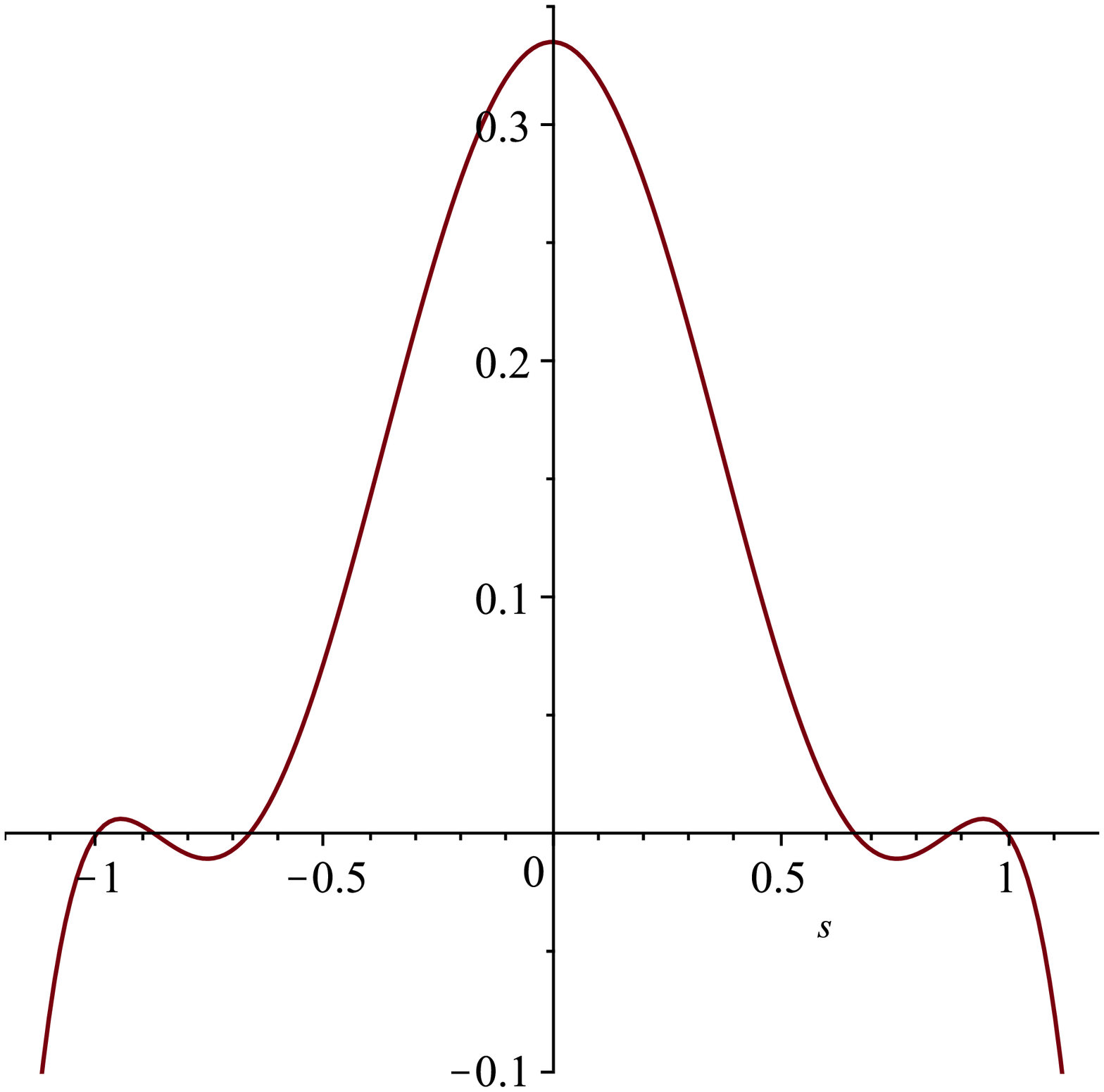} \hspace*{1cm}
\raisebox{3cm}{II}
\includegraphics[width=4cm]{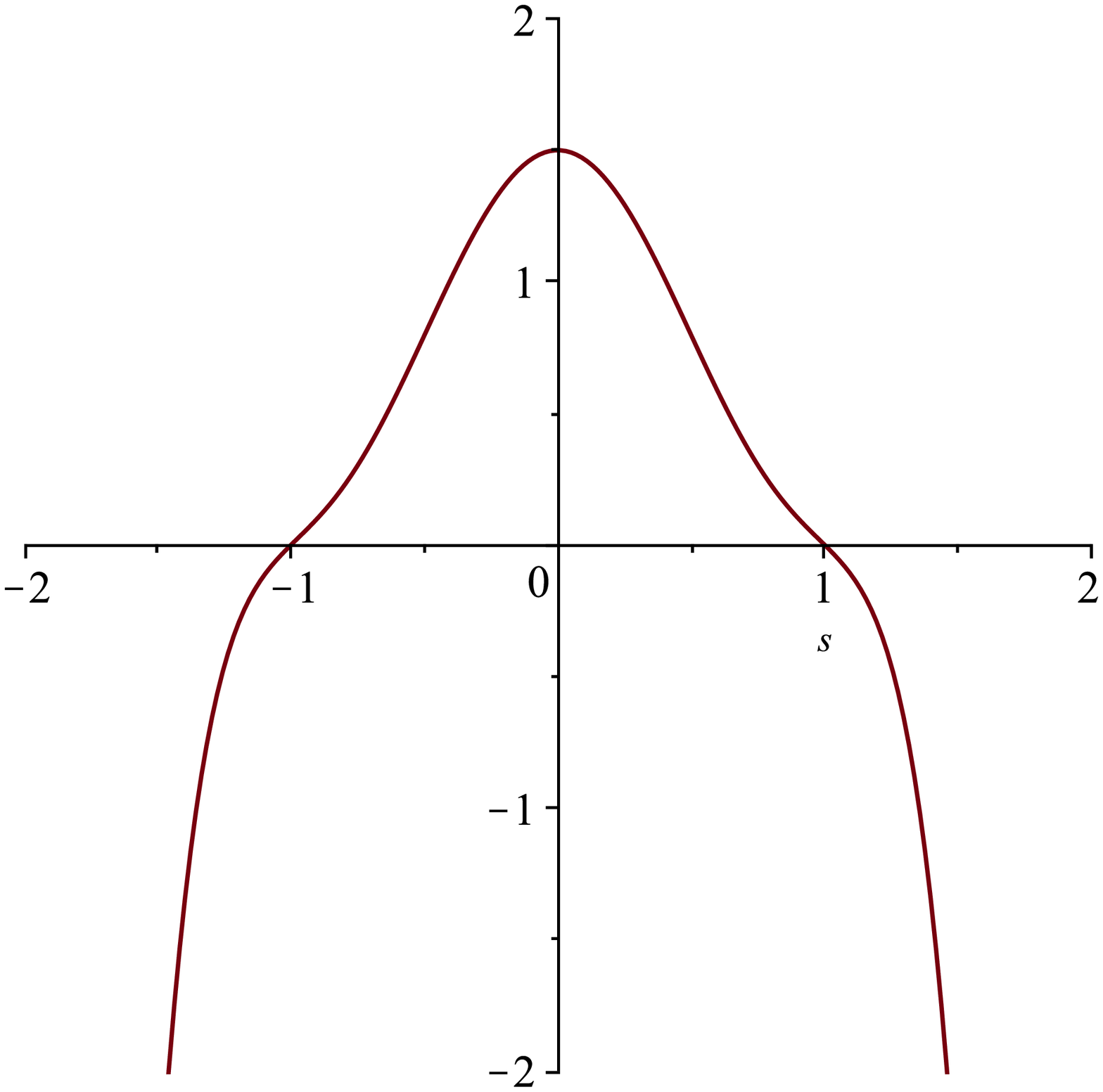} \\[1.5ex]
\raisebox{3cm}{III}
\includegraphics[width=4cm]{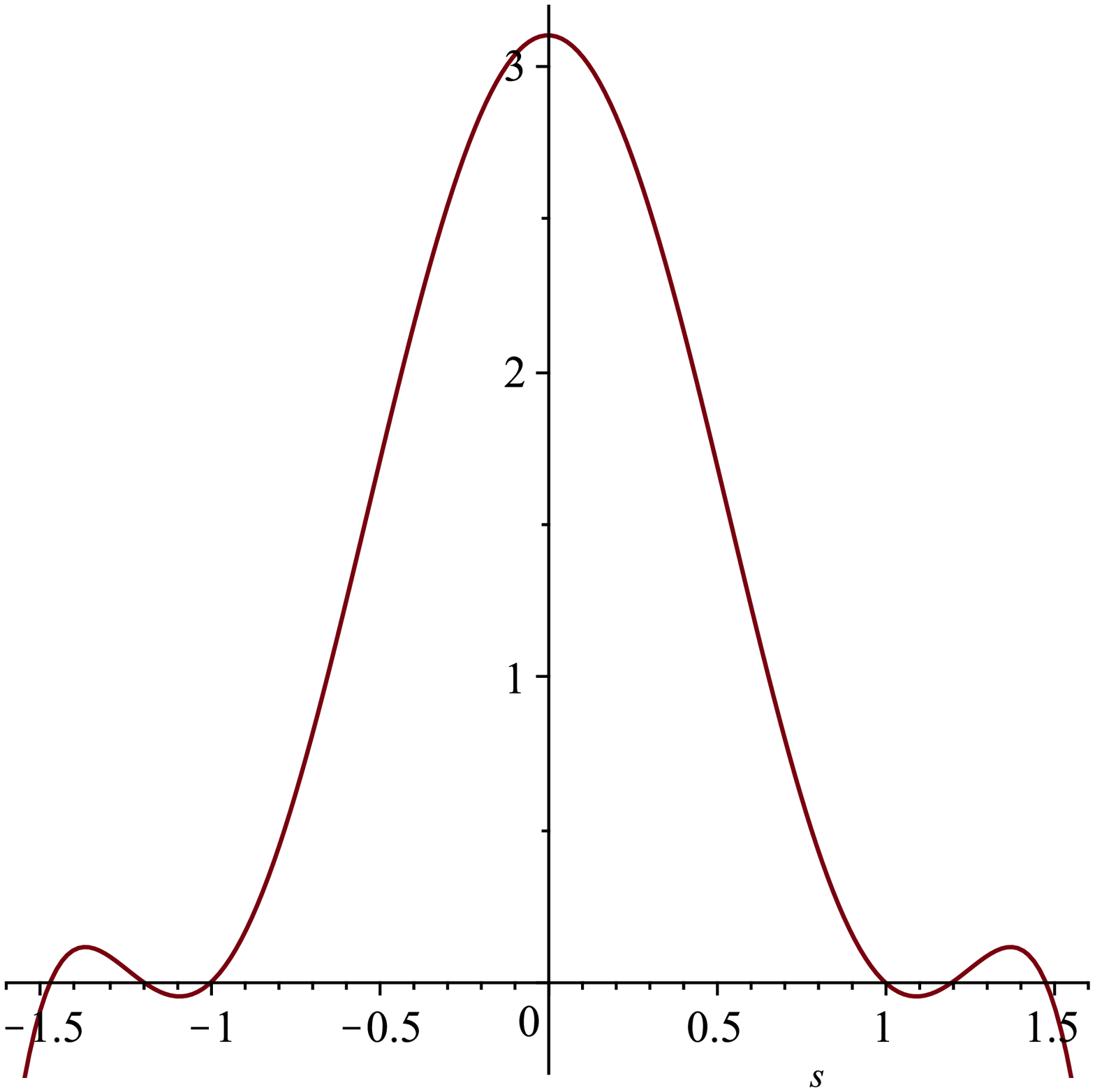}\hspace*{1cm}
\raisebox{3cm}{IV}
\includegraphics[width=4cm]{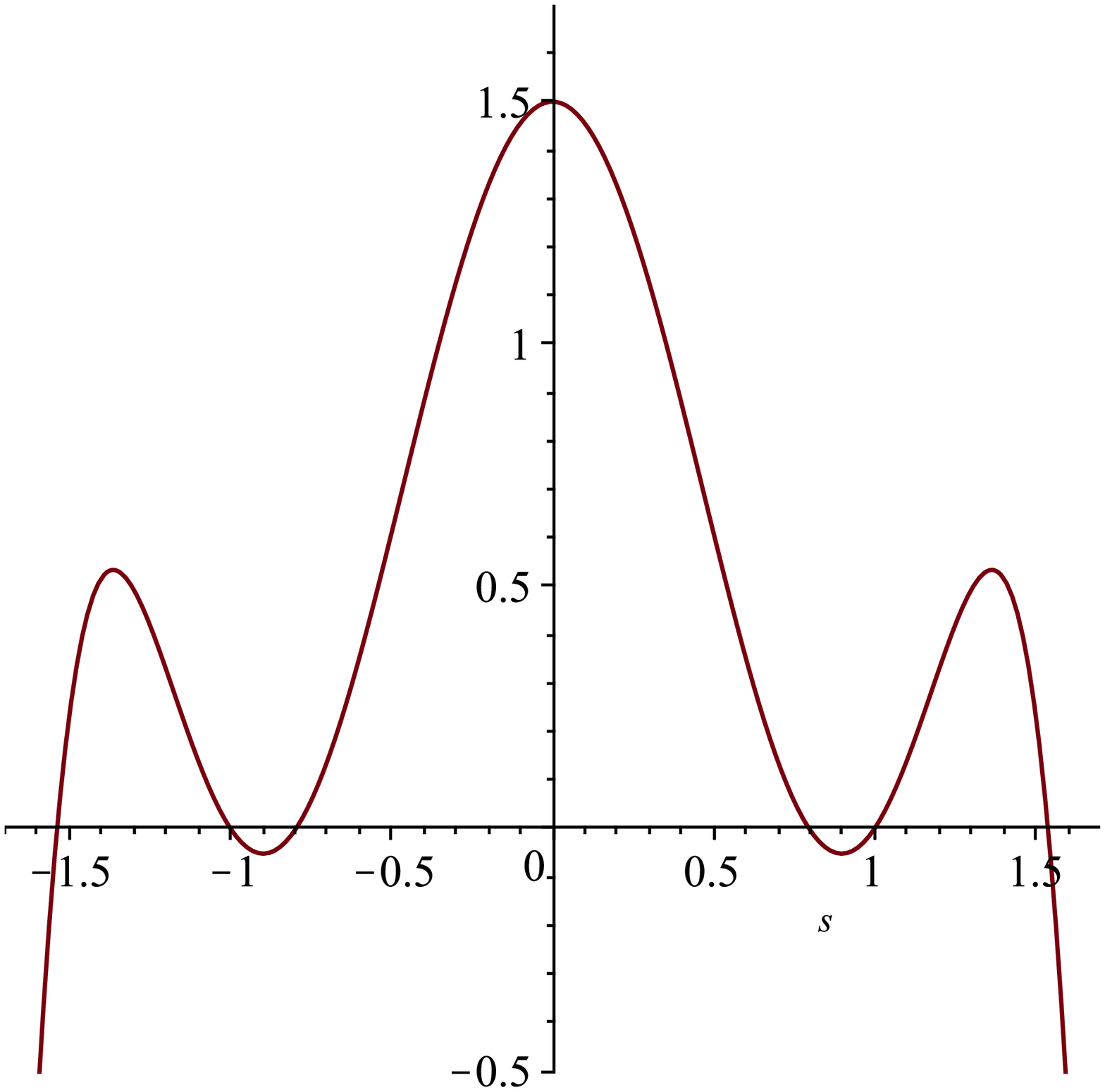} \hspace*{1cm}
\raisebox{3cm}{V}
\includegraphics[width=4cm]{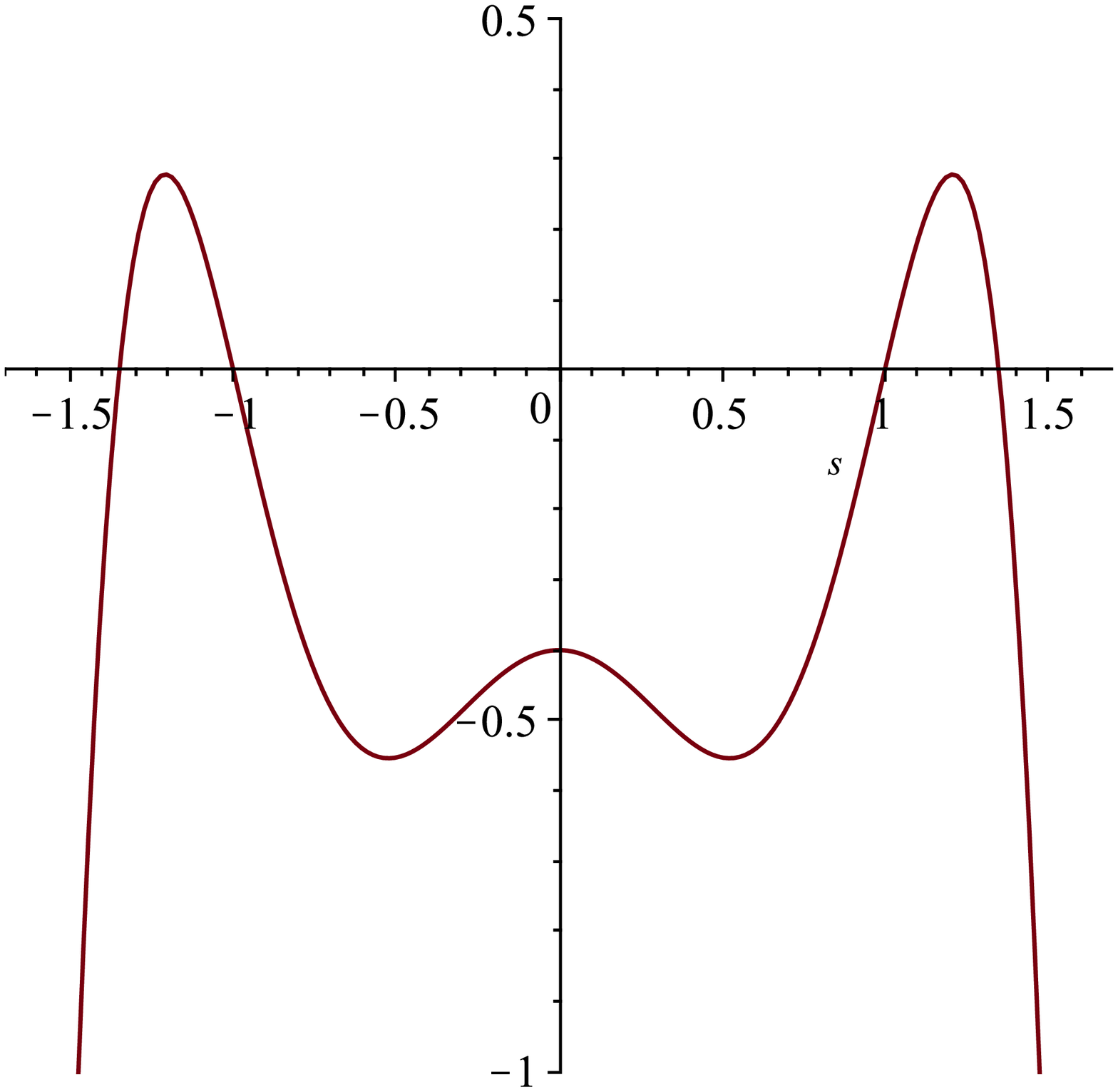}
\caption{Bifurcation diagram of the planar harmonic oscillator with energy momentum map $(H,G)$ (top left). The remaining panels show the graphs of the polynomial $P(s)$ for representative values of $(E,g)$ in the regions $I$ to $V$ marked in the $(g,h)$ plane. 
In region I: all roots are real and satisfy $|s_{2\pm}|<|s_{3\pm}|<|s_{1\pm}|$. 
In region II: $s_{2\pm}$ and $s_{3\pm}$ are complex.
In region III: all roots are real and satisfy $|s_{1\pm}|<|s_{2\pm}|<|s_{3\pm}|$. 
In region IV: all roots are real and satisfy $|s_{2\pm}|<|s_{1\pm}|<|s_{3\pm}|$. 
In region V: $s_{2\pm}$ are complex and $s_{3\pm}$ are real with $|s_{1\pm}|<|s_{3\pm}|$.
} \label{fig:bifdiag_planar}
\end{figure}

From the one-parameter family of two-dimensional harmonic oscillators with $l_z=0$ we will consider the one in the $(x,z)$ plane. This is an integrable system with the energy momentum map $(H,G)$ where $H$ and $G$ are the constants of motion defined in \eqref{eq:defH} and \eqref{eq:G_Cart} restricted to $y=p_y=0$.  
For $l_z = 0$, the roots of $P(s)$ are 
\begin{equation*} 
\begin{split}
s_{1\pm} &=\pm 1,  \\ 
s_{2\pm} &= \pm
{\frac{1}{2\omega a} {\sqrt { 2 {a}^{2}{\omega}^{2}+4\,h - 2\,
\sqrt{    (a^2\omega^2-2E)^2       -    4\, g  {\omega}^{2}}
}}} ,\\
  s_{3\pm} &= \pm 
{\frac{1}{2\omega a} {\sqrt { 2 {a}^{2}{\omega}^{2}+4\,h + 2\,
\sqrt{    (a^2\omega^2-2E)^2       -    4\, g  {\omega}^{2}}
}}} .
\end{split}
\end{equation*}
For values $(E,g)$ for which $s_{2\pm}$ and $s_{3\pm}$ are real, we have  $|s_{2\pm}| \le |s_{3\pm}|$.   If $s_{3\pm}$ are not real then $s_{2\pm}$ are also not real. But conversely $s_{3\pm}$ can be real even if $s_{2\pm}$ are not real.
The discriminant of $P(s)$ is 
$$ 
\text{discrim} (P(s),s) = 64\,  {a}^{12}{\omega}^{2}  
\left( 2\,{a}^{2}E+g \right) 
g^4
\left(   (a^2\omega^2-2E)^2    - 4\, g { \omega}^{2}    \right) ^{2} .
$$
Double roots occur for 
$$
{\cal L}_1 := \{ g=-2\,a^2 E  \}, \, 
{\cal L}_2 := \{ g= 0 \}, \, 
{\cal L}_3 := \{ g= {\frac {(a^2\omega^2-2\,E)^2    }{{
4\,\omega}^{2}}}
 \}.
$$
The curves ${\cal L}_i$, $i=1,2,3$, divide the upper $(g,E)$ half plane into five region with different dispositions of roots as shown in Fig.~\ref{fig:bifdiag_planar}. 
From the separated momenta in \eqref{eq:separated_momenta} we see that the values of the constants of motion facilitate 
physical motion (i.e. real momenta) if the resulting $P(s)$ is positive somewhere in $[-1,1]$ and at the same time positive somewhere in $[1,\infty)$. From Fig.~\ref{fig:bifdiag_planar} we see that this is the case only for regions III and IV.  
For a fixed energy $E\ge0$, the minimal  value of $g$ is determined by the collision of the roots $s_{2\pm}$ at $0$.
Whereas for a fixed energy $E> \frac12 \omega^2a^2$, the maximal value of $g$ is determined by the collision of the pairs of roots $s_{2\pm}$ and $s_{3\pm}$, the maximal value of $g$ 
for a fixed energy $0< E < \frac12 \omega^2a^2$ is determined by the collision of the pairs of roots  
$s_{3\pm}$ and $s_{1\pm}=\pm1$. At the boundary between regions III and IV, the pairs of roots $s_{2\pm}$ and $s_{1\pm}=\pm1$ collide.

\begin{figure}
\raisebox{4.5cm}{(a)} \includegraphics[height=4.3cm]{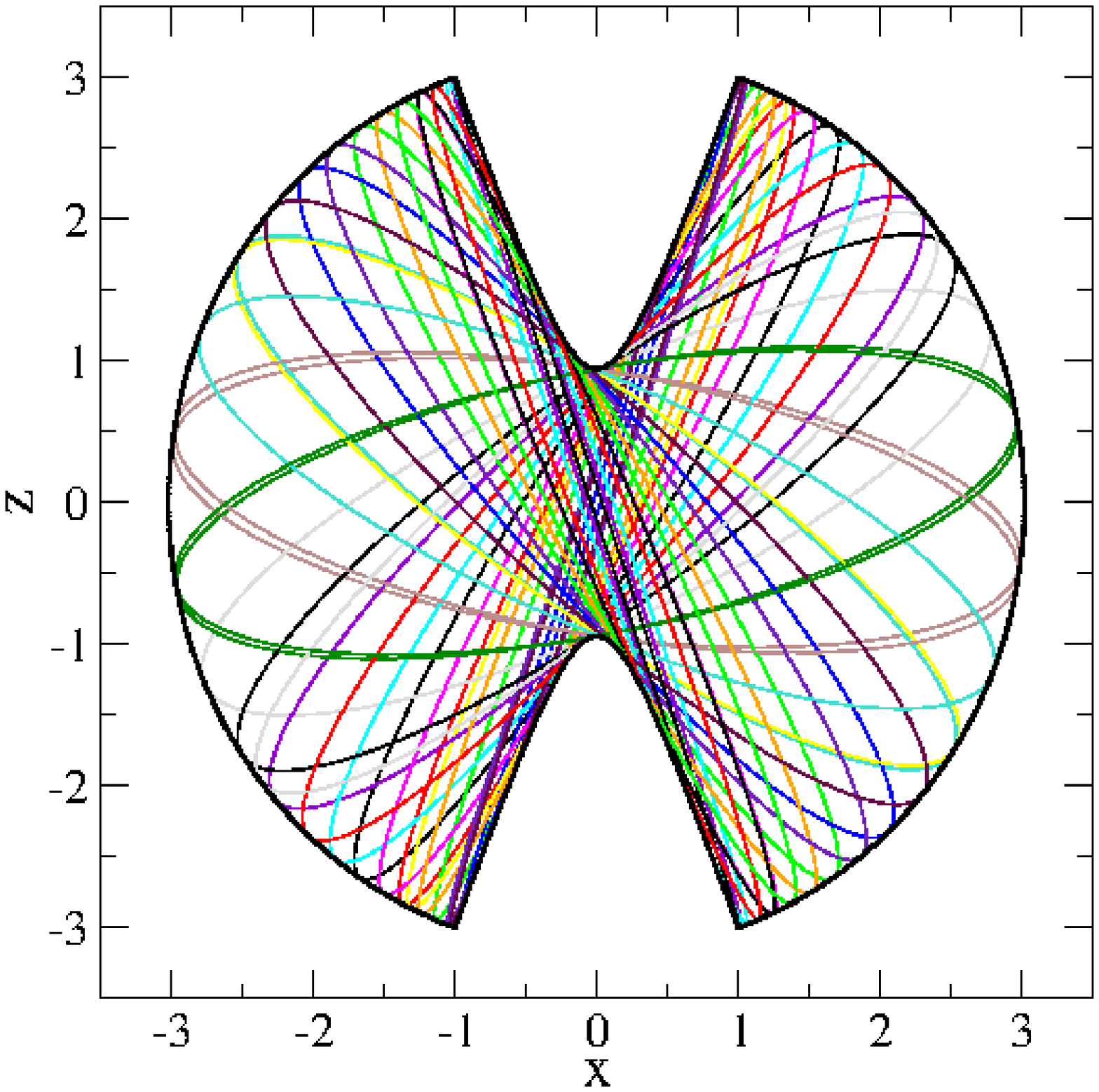}
\raisebox{4.5cm}{(b)} \includegraphics[height=4.3cm]{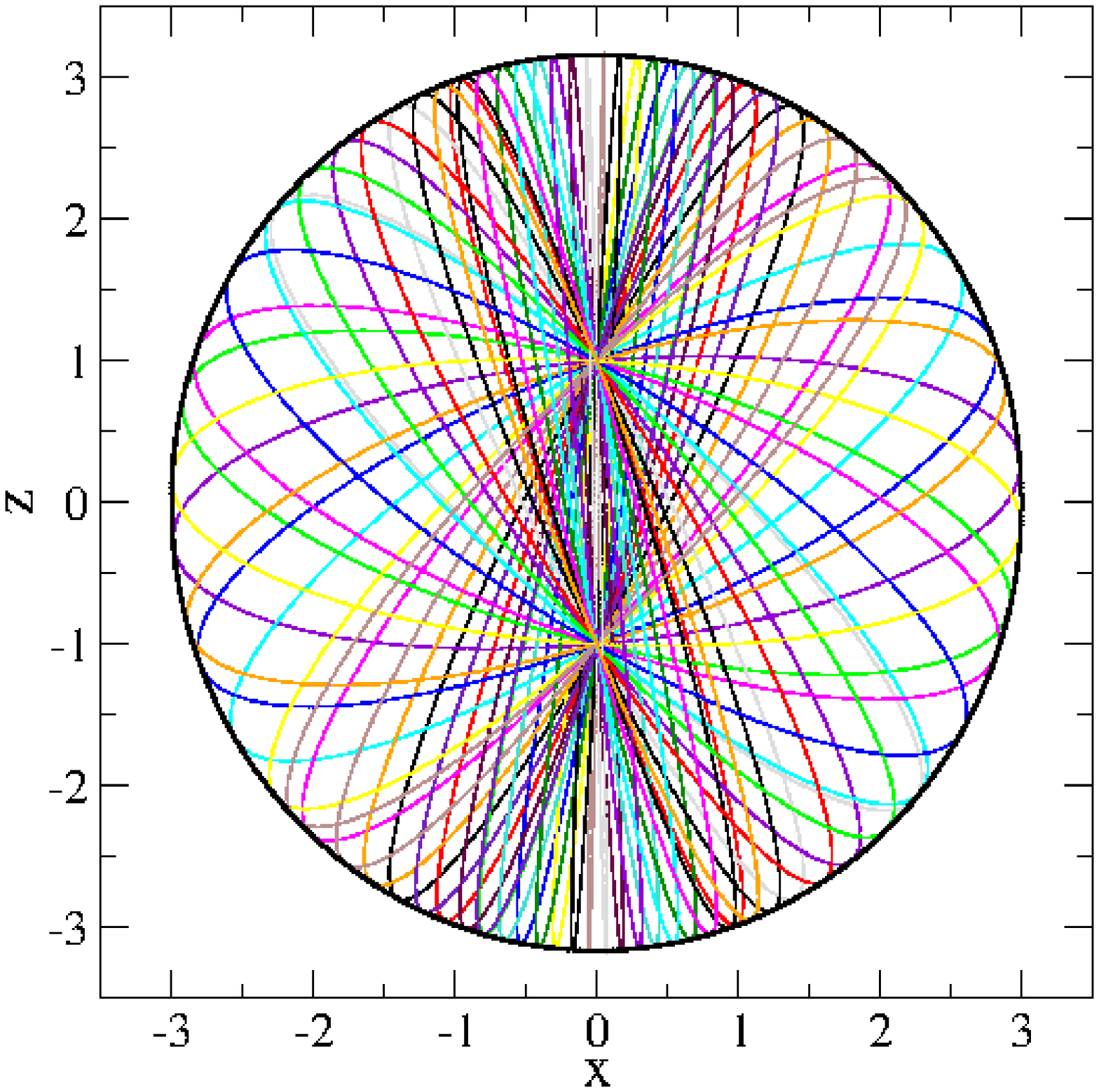}
\raisebox{4.5cm}{(c)} \includegraphics[height=4.3cm]{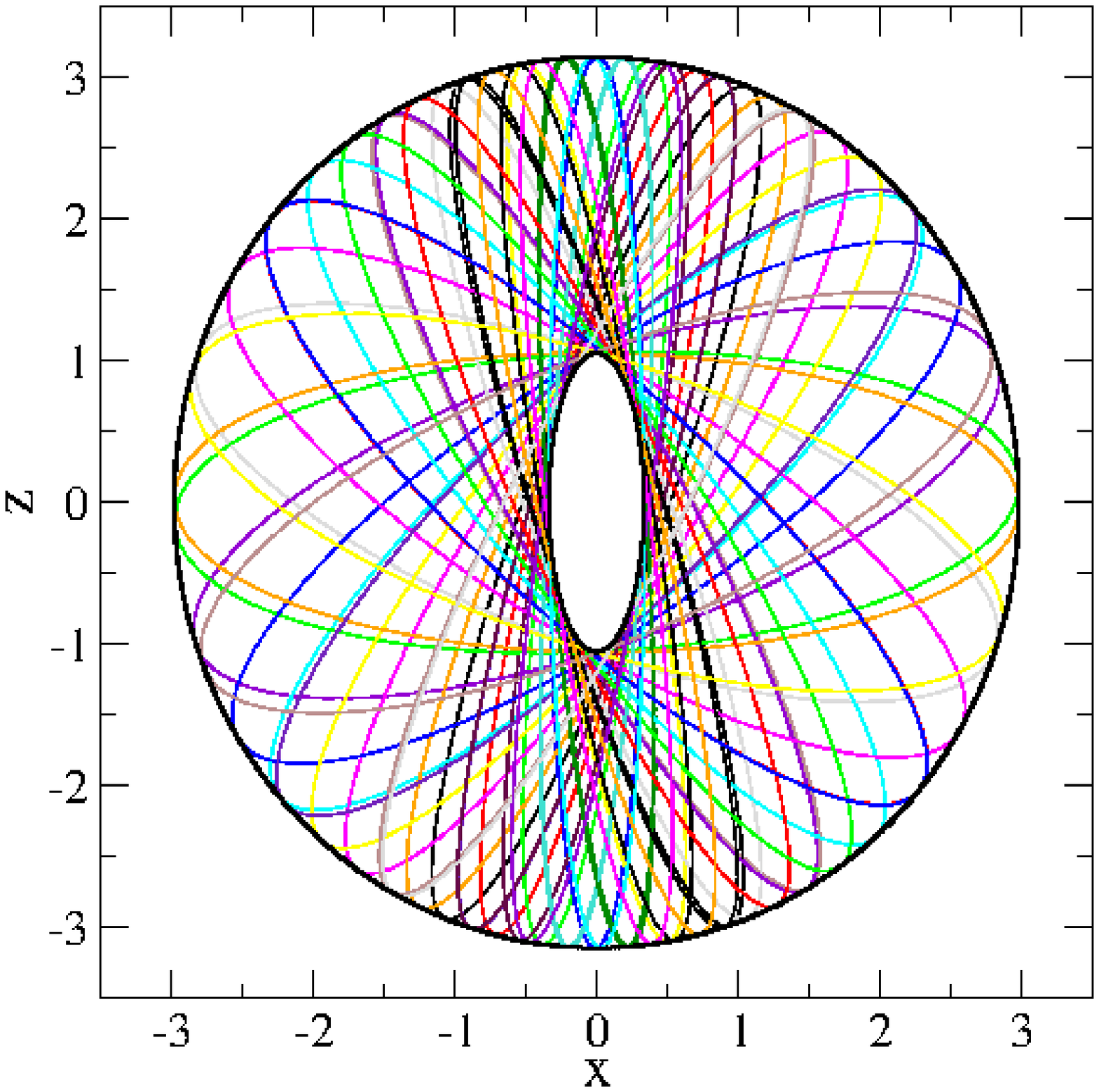}
\raisebox{4.5cm}{(d)} \includegraphics[height=4.3cm]{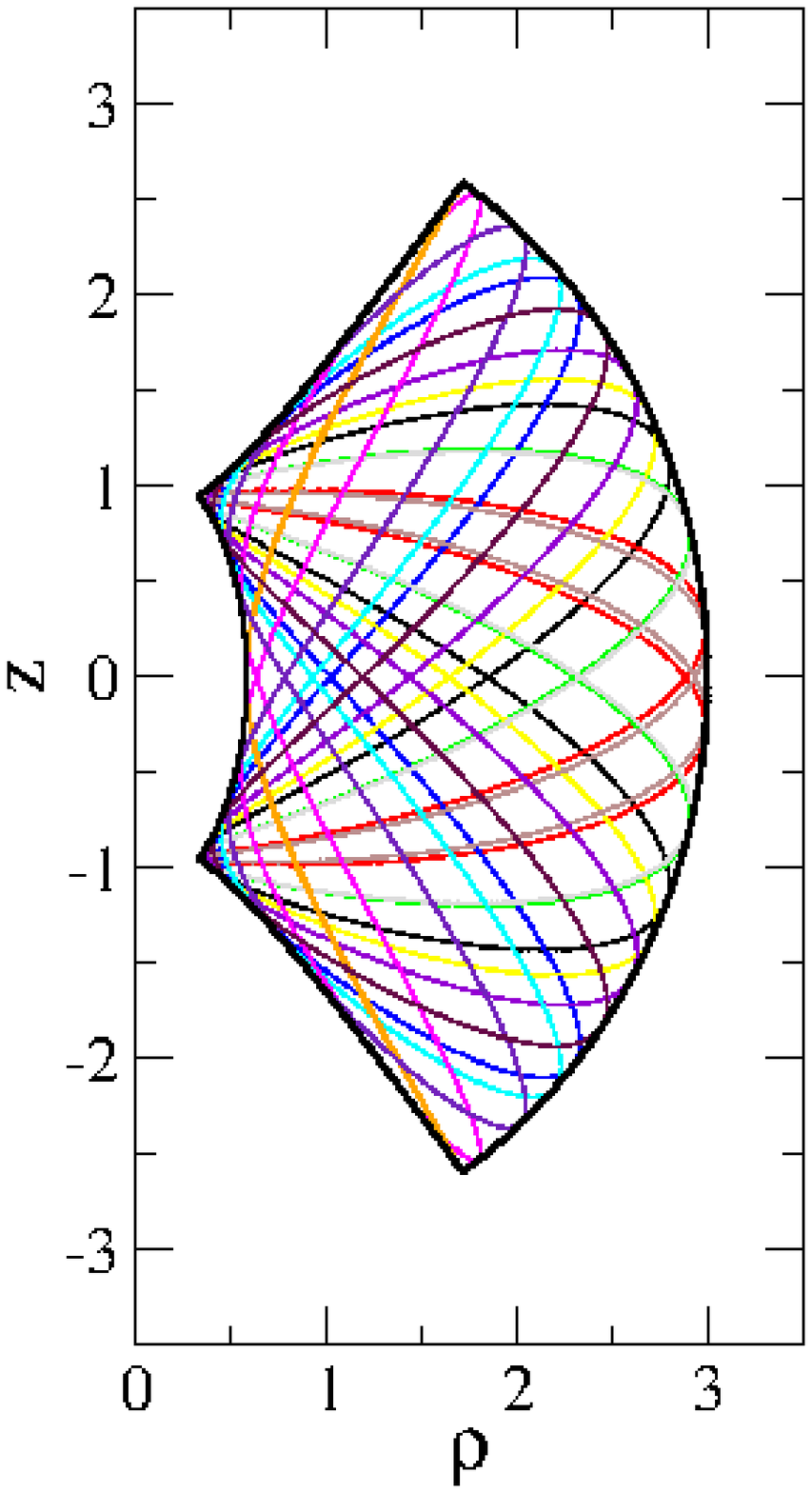}
\caption{Orbits and caustics for $h=5$, $l_z=0$ and $g=-1$ (region IV) in (a), $g=0$ (boundary III/IV) in (b) and $g=1$ (region III) in (c), and $h=5$, $l_z=1$ and $g=0$ in (d), where $\rho=\sqrt{x^2+y^2}$. In all panels $a=1$.
} \label{fig:caustics}
\end{figure}

For a value $(E,g)$ in region IV, the preimage under the energy momentum map $(H,G)$ is a two-torus consisting of a one-parameter family of periodic orbits whose projection to configuration space are ellipses which are enveloped by a caustic formed by the ellipse given by the coordinate line $\xi =  s_{3+}$ and the two branches of the confocal hyperbola corresponding to the coordinate line $\eta= s_{2+}$ (see Fig.~\ref{fig:caustics}a).  
For a value $(E,g)$ in region III, the preimage under the energy momentum map $(H,G)$ is a two-torus consisting of a one-parameter family of periodic orbits whose projection to configuration space are ellipses which are enveloped  by a caustic formed by two confocal ellipses given by the coordinate lines $\xi = s_{2+}$ and $\xi = s_{3+}$, respectively  (see Fig.~\ref{fig:caustics}c).
The boundary ${\cal L}_2 = \{ g=0 \}$ between regions III and IV  is formed by critical values of the energy momentum map $(H,G)$ and the preimage consists of a one-parameter family of periodic orbits whose projection to the configuration space are ellipses which each contain the focus points $\pm \mathbf{a}$ (see Fig.~\ref{fig:caustics}b). The family in particular contains the periodic orbit oscillating along the $z$ axis with turning points $z_\pm = \pm \sqrt{2E} / \omega$, where $|z_\pm|>a$. The caustic is again formed by the ellipse  $\xi =  s_{3+}$.  
For $(E,g)\in {\cal L}_2$ and $E<\frac12 \omega^2 a^2$, the preimage consists only of the periodic orbit oscillating along the $z$ axis between  $z_\pm = \pm \sqrt{2E} / \omega$ where $z_\pm $ now has a modulus less than $a$.  
For $(E,g)\in {\cal L}_3$, i.e. the maximal value of $g$ for fixed energy $E>\frac12 \omega^2 a^2$,
 the preimage consists of two periodic orbits whose configuration space projections are the ellipse $\xi=s_{2+}=s_{3+}$.  
For $(E,g)\in {\cal L}_1$, i.e. the minimal value of  $g$ for fixed energy $E$, the preimage consists of the periodic orbit that is oscillating along the $x$ axis with turning points  $x_\pm = \pm \sqrt{2E} / \omega$.  
The tangental intersection of  ${\cal L}_2$ and ${\cal L}_3$ at $(g,E)=(0,\frac12 \omega^2 a^2)$ corresponds to a pitchfork bifurcation where two ellipse shaped periodic orbits grow out of the periodic orbit along the $z$ axis.

\subsection{The three-dimensional harmonic oscillator (general $l_z$)}
\label{sec:bifdiag_gen}

\begin{figure}
\raisebox{4.5cm}{(a)}
\includegraphics[width=4.5cm]{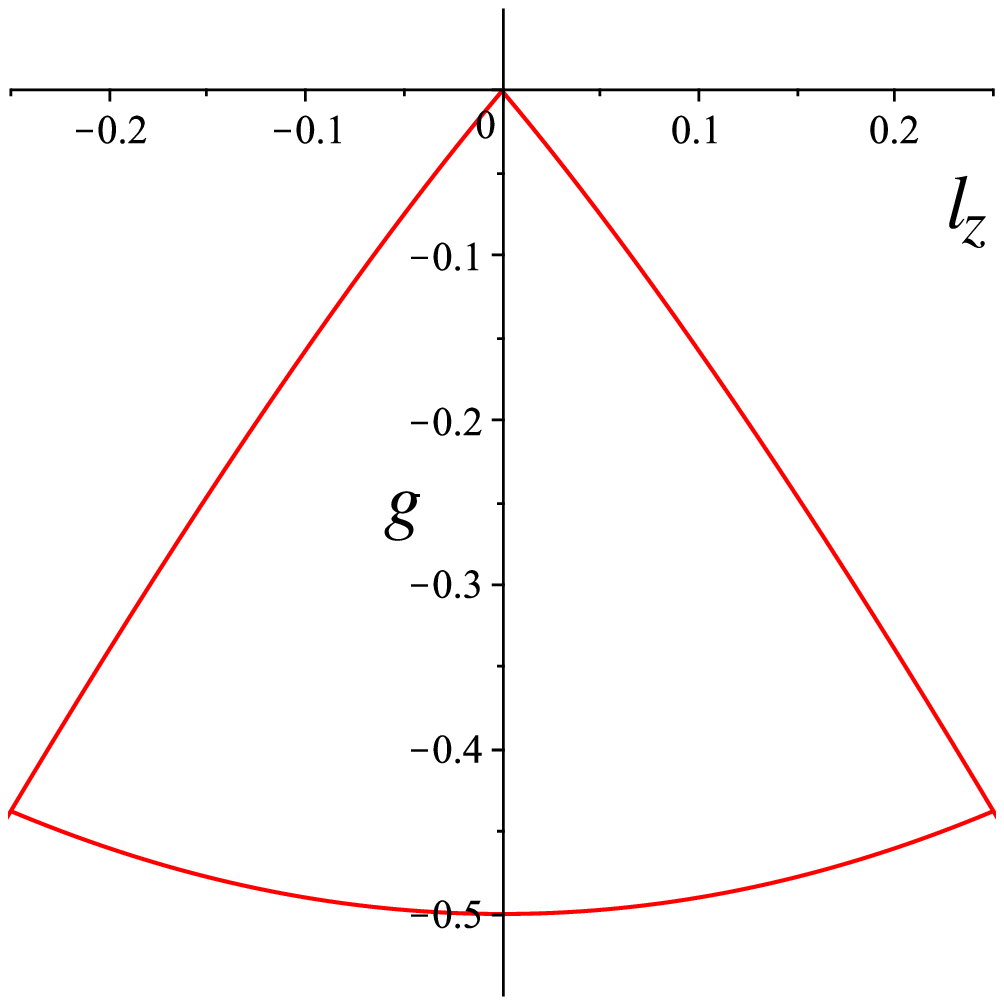} 
\raisebox{4.5cm}{(b)}
\includegraphics[width=4.5cm]{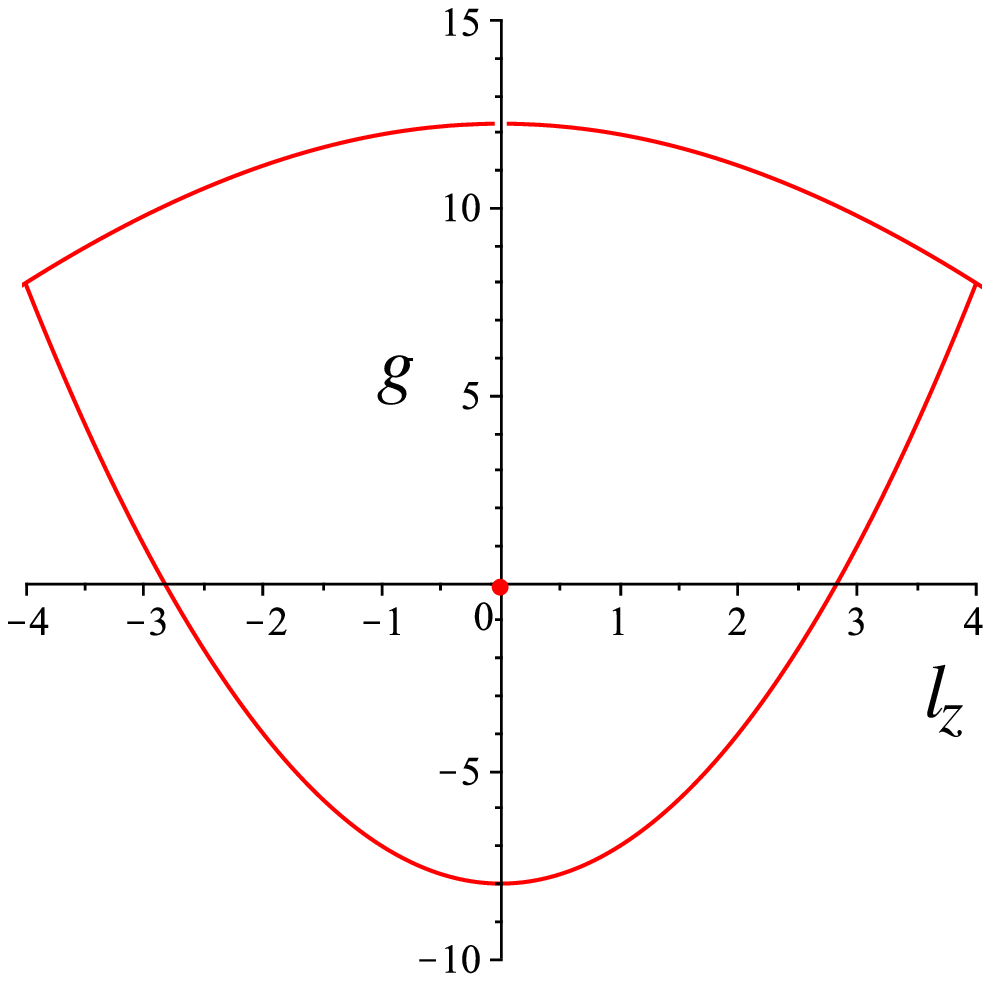} 
\caption{Slices of constant energy through the spatial bifurcation diagram with $a=1$, $\omega=1$ and energies $E=1/4$ (a) and $E=4$ (b).
} \label{fig:bifdiag_spatial}
\end{figure}

Increasing the modulus of $l_z$ from zero we see from the definition of $P(s)$ in Eq.~\eqref{eq:def_P(s)} 
that the graphs of the polynomial in Fig.~\ref{fig:bifdiag_planar} move downward. Even though we cannot easily give expressions for the roots of 
$P(s)$ for $l_z\ne 0$ we see that
increasing  $|l_z|$ from zero for fixed $E$ and $g$ the ranges of admissible $\eta$ and $\xi$ shrink. Moreover, as $P(\pm 1)=-l_z^2$,  
the roots  stay away from $\pm 1$ (the coordinate singularities of the prolate ellipsoidal coordinates) for $l_z \ne 0$. 
For general $l_z$, the discriminant of $P(s)$ is
\begin{equation*} 
\begin{split}
\text{discrim} (P(s),s) = 
64\, {a}^{12}{\omega}^{2} & \left( 2\,{a}^{2}E+g-{ l_z^2} \right) 
 \left( 4\,{a}^{8}{ l_z^2}\,{\omega}^{6}-24\,{a}^{6}E{ l_z^2}\,{
\omega}^{4}-{a}^{4}{g}^{2}{\omega}^{4}-18\,{a}^{4}g{ l_z^2}\,{
\omega}^{4}+ \right. \\ & \left. 
27\,{a}^{4}{{ l_z}}^4{\omega}^{4}+ 
48\,{a}^{4}{E}^{2
}{ l_z^2}\,{\omega}^{2}+4\,{a}^{2}{g}^{2}E{\omega}^{2}+36\,{a}^{2}g
E{ l_z^2}\,{\omega}^{2}-32\,{a}^{2}{E}^{3}{ l_z^2}+4\,{g}^{3}{
\omega}^{2}-4\,{g}^{2}{E}^{2} \right) ^{2}
\,.
\end{split}
\end{equation*}
%
The first (nonconstant) factor vanishes for
\begin{equation} \label{eq:min_g_general_lz} 
g =  {{\it l_z}}^{2} - 2\,{a}^{2}E. 
\end{equation}
From $P(0)=g-l_z^2+2\, a^2 E$ we see that this is the condition for the local maximum of $P(s)$ at $s=0$ to have the value zero or equivalently the collision of roots at 0. 
In order to see when the second nonconstant factor vanishes it is useful to write $P(s)$ as $(s-d)^2(a_4s^4 + a_3s^3 a_2s^2 + a_1s + a_0)$ where $d$ is the position of the double root. Comparing coefficients then gives
\begin{eqnarray}  
g(d) &=& -a^2 \left( {d}^{2}-1 \right)  \left({a}^{2}\omega^2  (3d^2-1) -4\,E \right) , \label{eq:g(d)} \\ 
l_z^2(d) &=& \phantom{-} a^2 \left( {d}^{2}-1 \right)^2 \left( {{a}^{2}\omega^2  (2d^2-1) -2\,E} \right) .  \label{eq:lz(d)}
\end{eqnarray}
For fixed $E$ and $l_z$, the minimal value of $g$ is, similarly to the planar case ($l_z=0$),
determined by the occurrence of a double root of $P(s)$ at $0$, i.e. by Eq.~\eqref{eq:min_g_general_lz}.
The maximal value of $g$ for fixed $E$ and $l_z$ is similarly to the planar case determined by the collision of the two biggest roots of $P(s)$ and given by $g(d)$ in Eq.~ \eqref{eq:g(d)} for the corresponding $d>1$. 
We present the bifurcation diagram as slices of constant energy for representative values of $E$.
We have to distinguish between the two cases  $0<E<\frac12 \omega^2 a^2$ and  $E>\frac12 \omega^2 a^2$ as shown in Fig.~\ref{fig:bifdiag_spatial}. 
The upper branches of the bifurcation diagrams in Fig.~\ref{fig:bifdiag_spatial} result from $d>1$ in Eqs.~\eqref{eq:g(d)} and \eqref{eq:lz(d)}.
A kink at $l_z=0$ occurs when $E<\frac12 \omega^2 a^2$. This is because the second factor in  $\eqref{eq:lz(d)}$ can be zero at a $d\ge 1$ only if $E>\frac12 \omega^2 a^2$ in which case there is no kink. 
For  $E<\frac12 \omega^2 a^2$,  there is an isolated point at $(l_z,g)=(0,0)$.
This results from $d=\pm1$ in Eqs.~\eqref{eq:g(d)} and \eqref{eq:lz(d)}. The point is isolated because the second factor in  $\eqref{eq:lz(d)}$ is negative for  
$E>\frac12 \omega^2 a^2$ and $d=\pm 1$.
The preimage of a regular value of $(H,L_z,G)$  in the region enclosed by the outer lines bifurcation diagrams in Fig.~\ref{fig:bifdiag_spatial} corresponds to a three-torus formed by a two-parameter family of periodic orbits given by ellipses in configuration space which are enveloped by two-sheeted hyperboloids and two ellipsoids given by coordinate surfaces of the prolate spheroidal coordinates $\eta$ and $\xi$, respectively (see Fig.~\ref{fig:caustics}d).
The preimage of a critical value $(E,l_z,g)$ in the upper branches in Fig.~\ref{fig:bifdiag_spatial} is a two-dimensional torus consisting of periodic orbits that move on ellipsoids of constant $\xi$. 
The preimage of a critical value $(E,l_z,g)$ in the lower branches consists of a two-dimensional torus formed by periodic orbits whose projections to configuration space are contained in the $(x,y)$ plane. 
At the corners where $|l_z|$ reaches its maximal value $E/\omega$, the motion is along the circle of radius $(|l_z|/\omega)^{1/2}$ in the $(x,y)$ plane with the sense of rotation being determined by the sign of $l_z$.


For the planar case, we saw that the critical energy $E=\frac12 \omega^2 a^2$ corresponds to a pitchfork bifurcation. In the spatial case this becomes a Hamiltonian Hopf bifurcation which manifests itself as the vanishing of the kink and detachment of the isolated point in the bifurcation diagram when $E$ crosses the value $\frac12 \omega^2 a^2$. Note that the critical energy is the potential energy at the focus points of the prolate spheroidal coordinates.


\subsection{Reduction}
\label{sec:reduction}

The isolated point of the bifurcation diagram for energies $E>\frac12 \omega^2 a^2 $ leads to monodromy.  
To see this more rigorously we proceed as follows.
For a classical maximally super-integrable Hamiltonian with compact energy surface, the flow of the Hamiltonian 
is periodic. Therefore it is natural to consider symplectic reduction by the $S^1$ symmetry induced by 
the Hamiltonian flow. This leads to a reduced system on a compact symplectic manifold.
On the reduced space which turns out to be $\mathbb{C}P^2$ we then have a two-degree-of-freedom Liouville integrable system $(L_z,G)$. 
We will prove  that for $E > \frac12 \omega^2 a^2$, this system has monodromy by showing the existence of a singular fibre with value $(l_z,g)=(0,0)$ (the isolated point discussed in the previous subsection) given by a 2-torus that is pinched at a focus-focus singular point. 
To this end it is useful to also reduce the $S^1$ action 
corresponding to the flow of $L_z$. As this $S^1$ action has isotropy, standard symplectic reduction is not applicable and we resort to singular reduction using the method of invariants instead. The result will be a one-degree-freedom system on a singular phase space. For a general introduction, we refer to \cite{CushBates}.

In order to reduce by the flows of $H$ and $L_z$ it is useful to rewrite $G$ as
\begin{equation}\label{eq:G_reduced}
G =  L_z^2 - 2 R^2  - \frac{2}{\omega}(a^2 \omega^2 - H) R +\frac{1}{ \omega} X, 
\end{equation}
where 
\begin{eqnarray} 
R &:=& \frac{1}{\omega}(A_x + A_y), \label{eq:def_R} \\
X &:=& \omega(L_x^2 + L_y^2) - 2A_z R. \label{eq:def_X}
\end{eqnarray}
The significance of this decomposition is that defining $Y$ by $$\{ R, X \} = -2  Y$$ we find 
that the Poisson brackets between $R$, $X$, and $Y$ are closed. Specifically we have
$$\{ R, Y \} = 2  X \text{ and }\{ X, Y \} = 8 (H - \omega R) (\omega L_z^2 + H R - 2 \omega R^2)$$
and $(R$,X$, Y)$ form a closed Poisson algebra  
with Casimir function
\begin{equation}\label{eq:DefCasimirFunction}
C=4 \omega^2  (H - \omega R)^2 ( R^2 - L_z^2) - \omega^2 ( X^2 + Y^2 ) = 0.
\end{equation}
Hence this achieves reduction to a single degree of freedom with phase space given by the zero level set of the Casimir function $C$. 

A systematic way to achieve this reduction uses invariant polynomials.
This approach is moreover useful because it gives a classical analogue to creation and annihilation operators used in the quantization below.
The flows generated by $(H, L_z)$ define a $T^2$ action on the original phase space $T^*\mathbb{R}^3$.
Since both $H$ and $L_z$ are quadratic and they satisfy $\{ H, L_z \}  = 0$ there is a linear symplectic transformation 
that diagonalises both $H$ and $L_z$. It is given by
\[
x = \frac{1}{\sqrt{2\omega}} (p_1+p_2),\ 
y = \frac{1}{\sqrt{2\omega}} (q_1-q_2),\ 
z = \frac{1}{\sqrt{\omega}} q_3,\ 
p_x = -\sqrt{\frac{\omega}{2}} (q_1+q_2),\ 
p_y = \sqrt{\frac{\omega}{2}} (p_1-p_2),\ 
p_z = \sqrt{\omega} p_3.
\]
and in the new complex coordinates $z_k = p_k + i q_k$, $k=1,2,3$, we find 
\[
    H = \frac{\omega}{2} ( z_1 \bar z_1 + z_2 \bar z_2 + z_3 \bar z_3), \quad
    L_z = \frac12( z_1 \bar z_1 - z_2 \bar z_2) \,.
\]
Additional invariant polynomials are
\[
    R = \frac12( z_1 \bar z_1 + z_2 \bar z_2), \quad
    X - i Y = \omega z_1 z_2 \bar z_3^2 \,.
\]
These invariants are related by the syzygy $C=0$ in Eq.~\eqref{eq:DefCasimirFunction}
and satisfy $|L_z| \le R \le H/\omega$.

\begin{figure}
\raisebox{4cm}{(a)}\includegraphics[width=5cm]{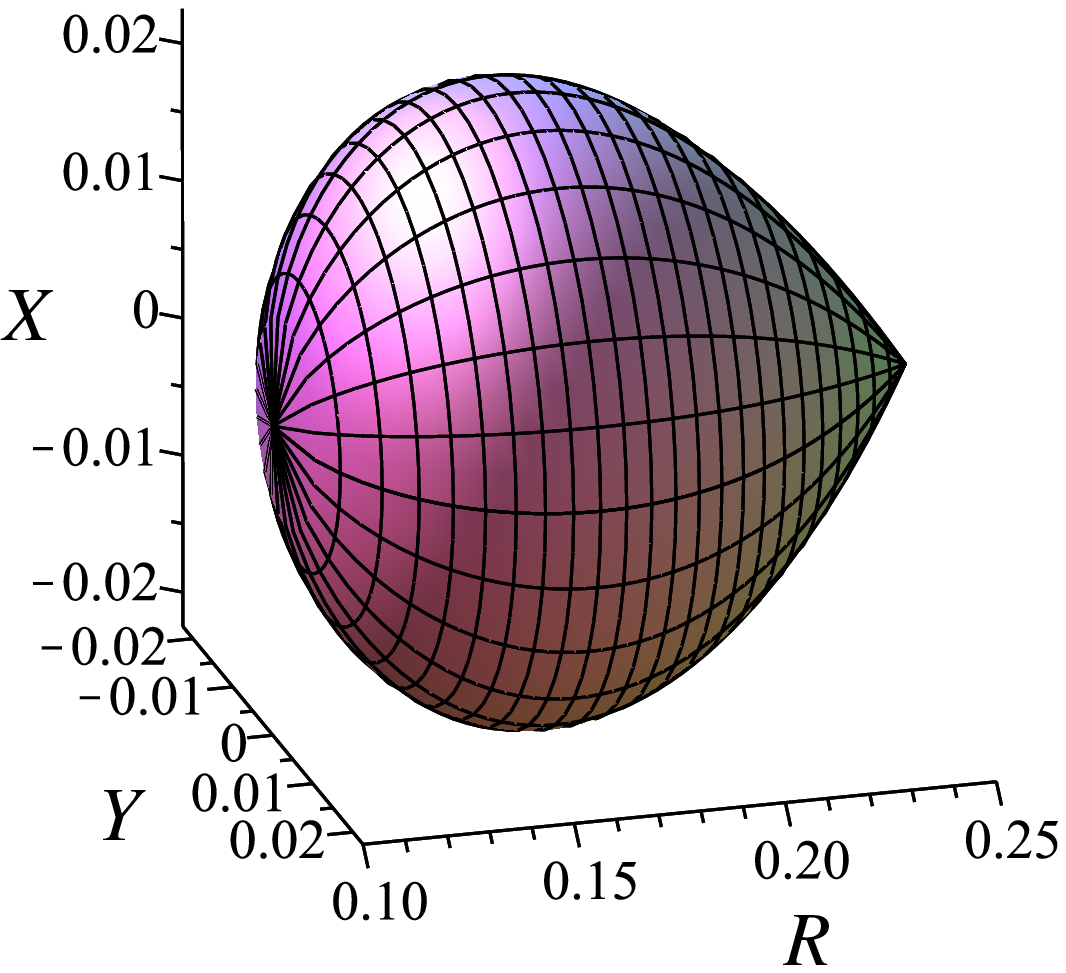}
\raisebox{4cm}{(b)}\includegraphics[width=5cm]{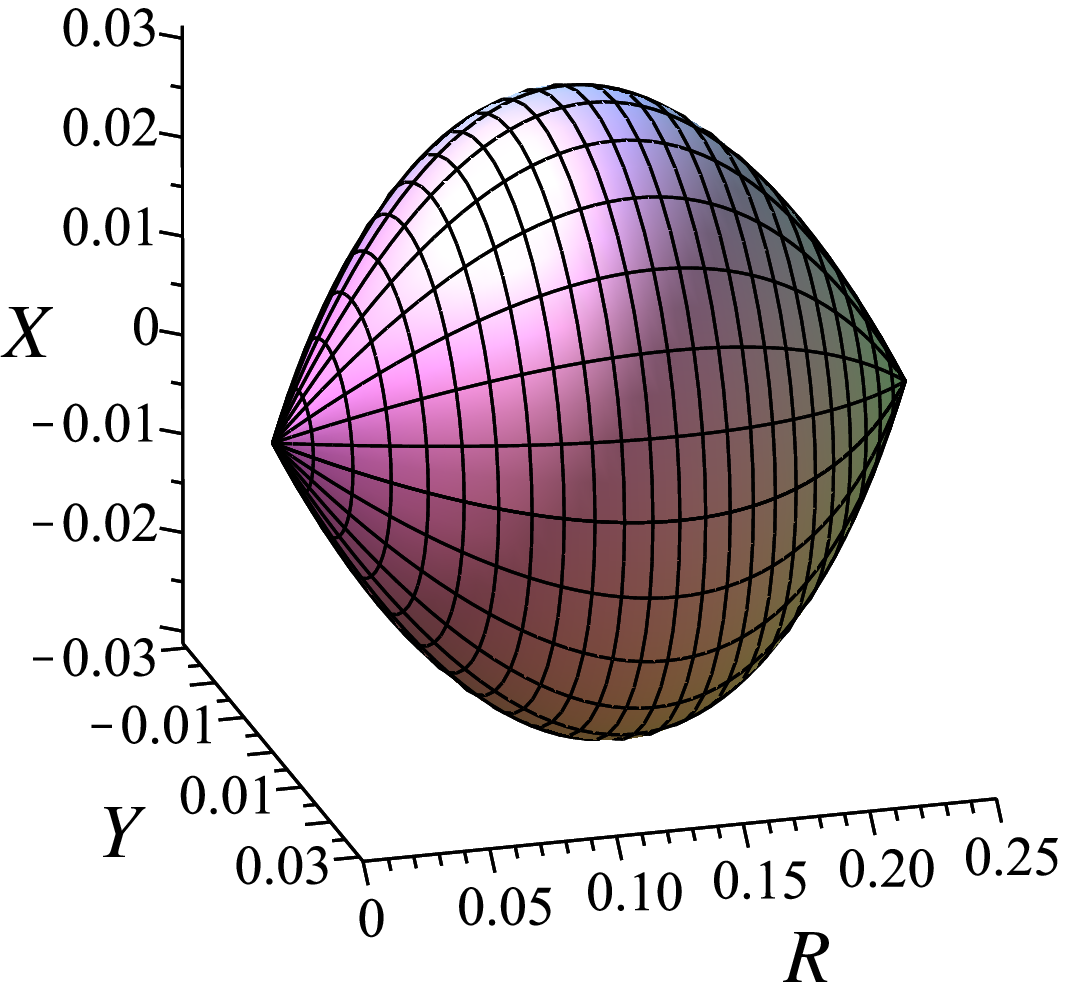}
\raisebox{4cm}{(c)}\includegraphics[width=5cm]{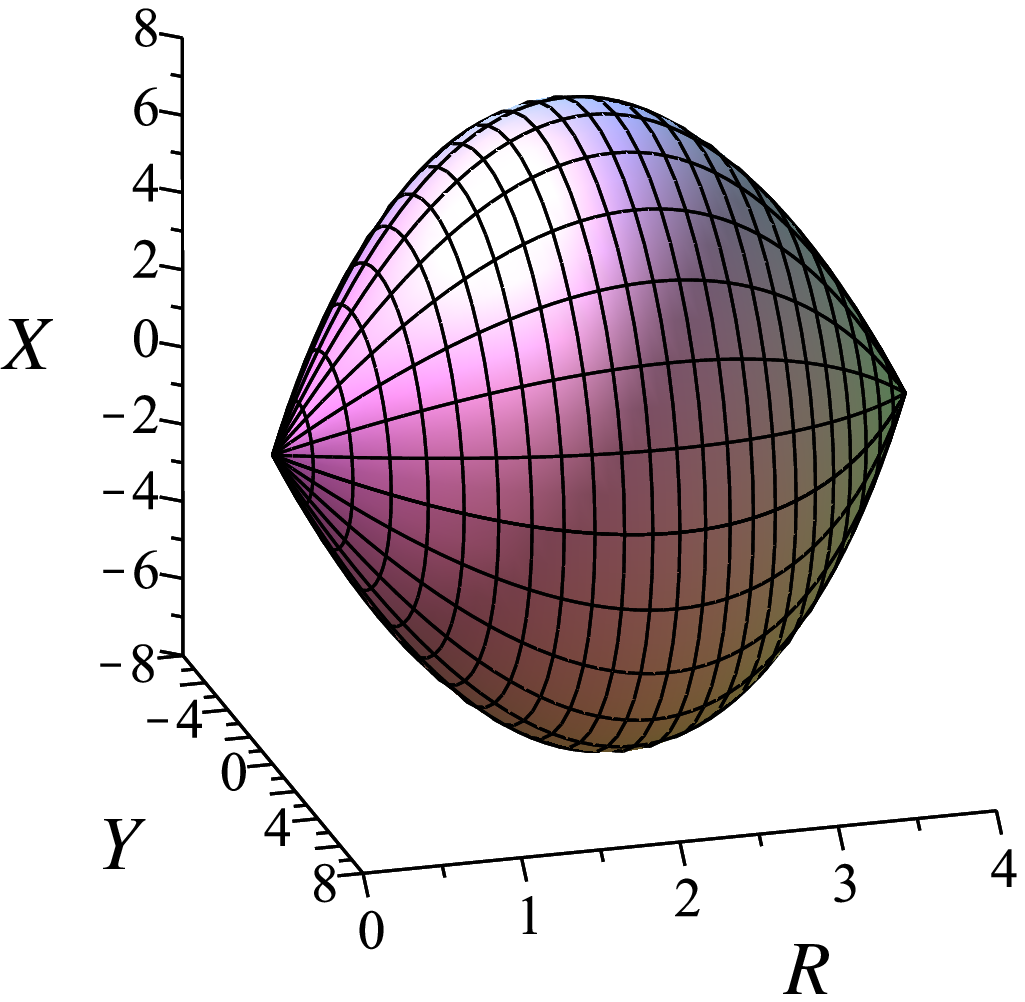} \\
\raisebox{5cm}{(d)}\includegraphics[width=5cm]{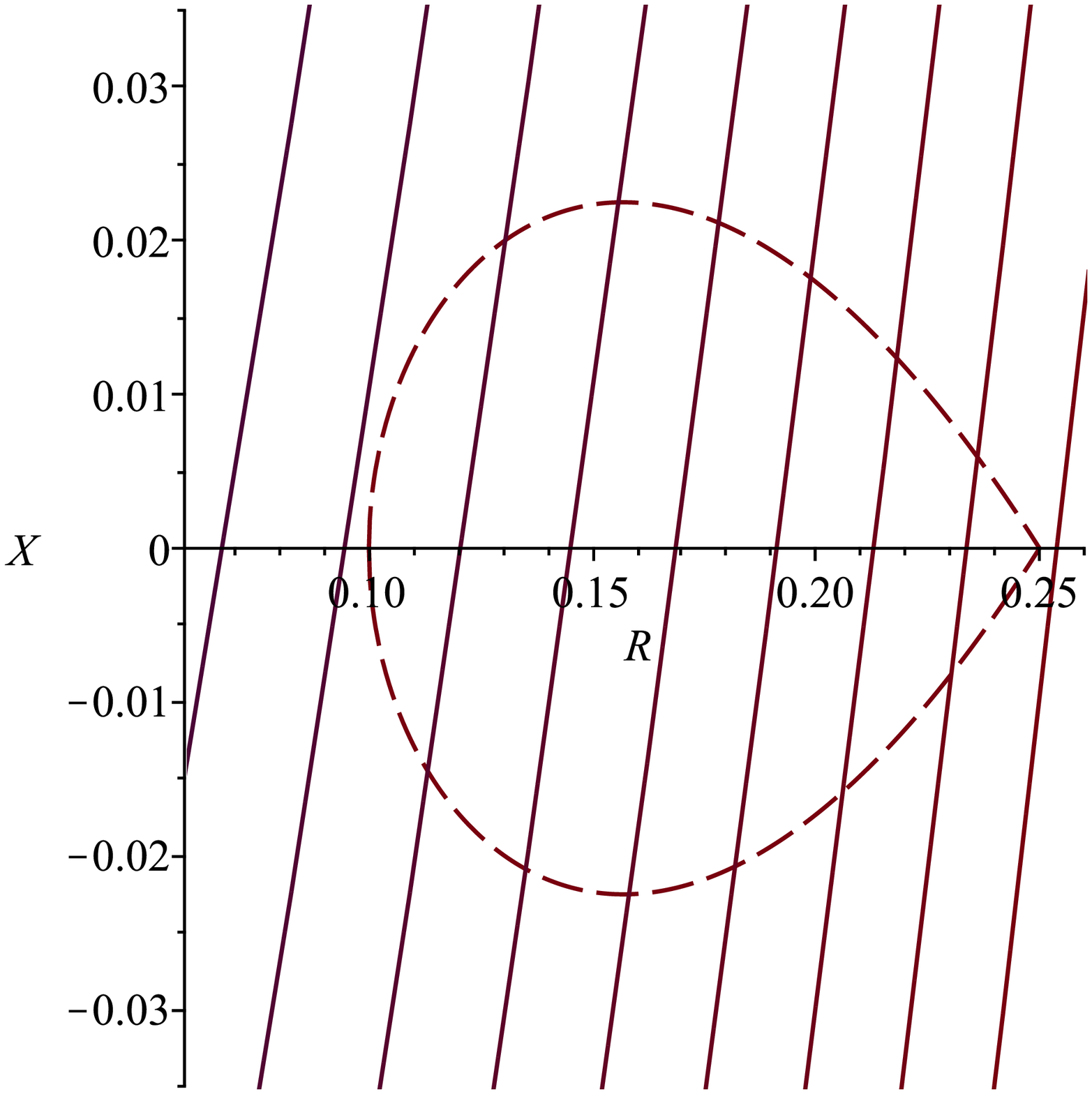}
\raisebox{5cm}{(e)}\includegraphics[width=5cm]{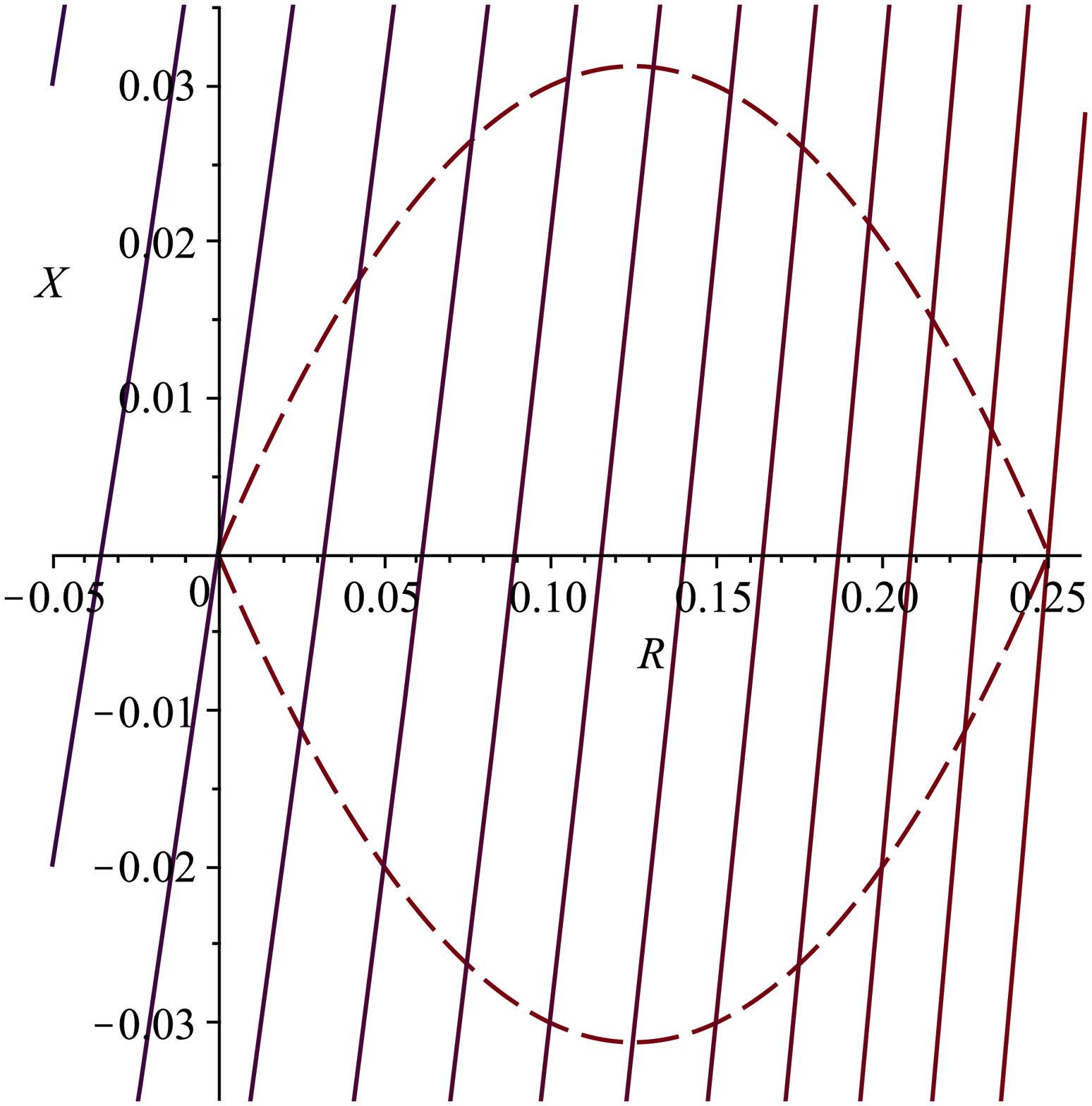}
\raisebox{5cm}{(f)}\includegraphics[width=5cm]{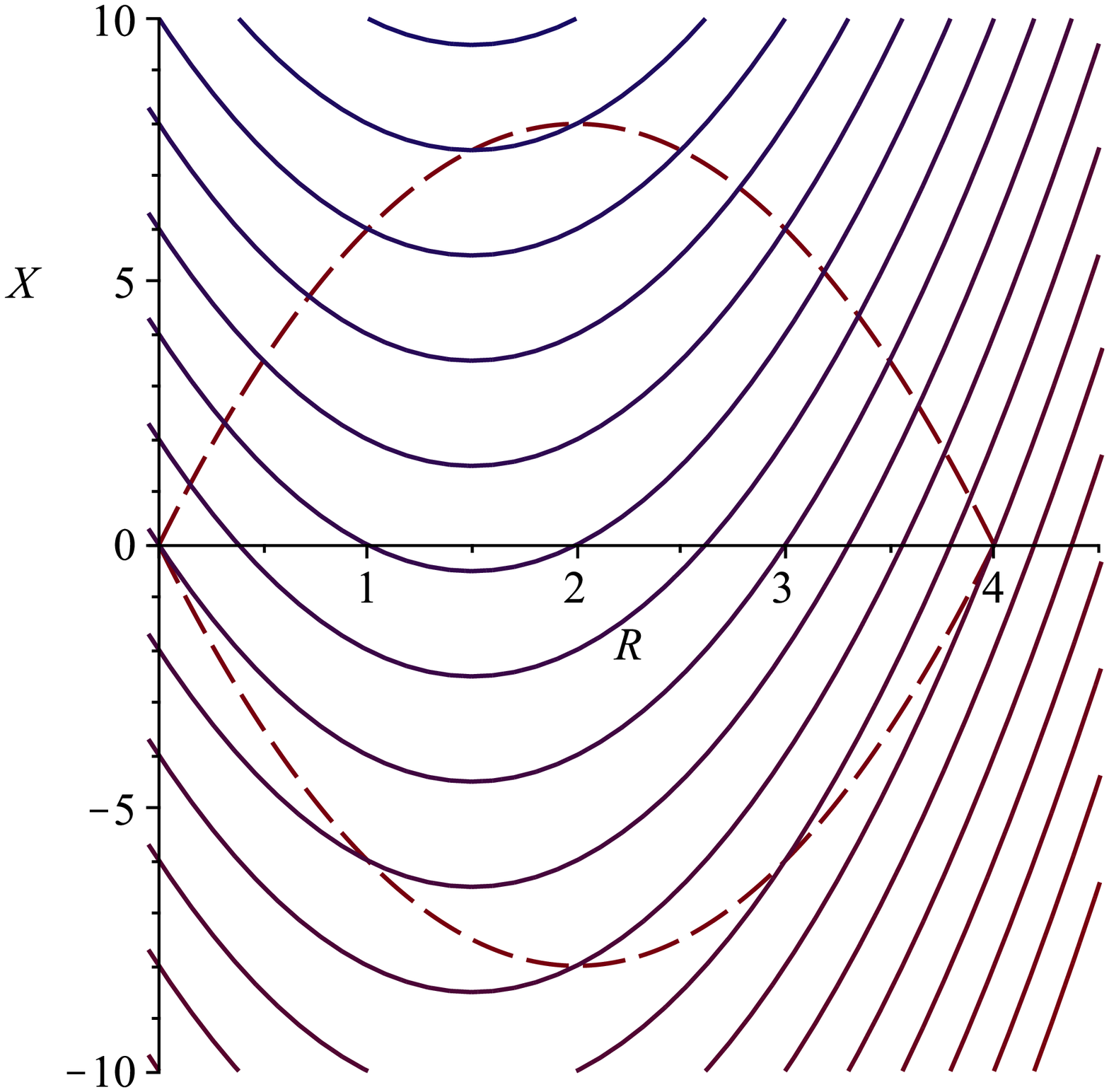} 
\caption{Reduced space $C=0$ for $E=1/4$ and $l_z=0.1$ (a),  $E=1/4$ and $l_z=0$ (b), and  $E=4$ and $l_z=0$ (c). 
The lower panels show the corresponding slices $Y=0$  (dashed) and contours $G=g$ with increments $\Delta g= 0.05$ in (d) and (e) and $\Delta g= 2$ in (f).
In all panels $a=\omega=1$.
} \label{fig:RedSpace}
\end{figure}

The surface $C=0$ in the three-dimensional space $(X,Y,R)$ can be viewed as the reduced phase space.  It is  rotationally symmetric about the 
$R$ axis. Due to a singularity at $R=E/\omega $ and another singularity at $R=0$ when $l_z=0$, the reduced space is
 homeomorphic  but not diffeomorphic to a  two-dimensional sphere (see Figs.~\ref{fig:RedSpace}(a) and (b)).
The singularity at $R=0$ when $l_z=0$ results from nontrivial isotropy of the $S^1$ action of the flow of $L_z$.  
$R=0$ implies that the full energy is contained in the $z$ degree of freedom and motion consists of oscillation along the $z$ axis. The corresponding phase space points are fixed points of the $S^1$ action of the flow of $L_z$. The value of $L_z$ is zero for this motion. For $R=E/\omega$,  the energy is contained completely in the $x$ and $y$ degrees of freedom (see \eqref{eq:def_R}), i.e. the motion takes place in the $(x,y)$ plane. This includes also the motion  along the circle of radius $(|l_z|/\omega)^{1/2}$ where  the flows of $L_z$ and $G$ are parallel. 

The dynamics on the reduced phase space is generated by $G$. As the system has only one degree of freedom
the solutions are given by the level sets of $G$ restricted to $C=0$.  
As $G$ is independent of $Y$ the surfaces of constant $G$ are cylindrical in the space $(X,Y,R)$. Given the rotational symmetry of the reduced phase space the intersections of $G=g$ and $C=0$ can  be studied in the slice $Y=0$ (see Fig.~\ref{fig:RedSpace}). Two intersection points in the slice result in a topological circle. Under variation of the value of the level $g$ the two intersection points collide at a tangency or the singular point where $R=E/\omega$ corresponding to the maximal and minimal values of $g$ for which there is an intersection, respectively. Both cases correspond to  elliptic equilibrium points for the flow of $G$ on the reduced space. For $l_z=0$, one of the intersection points can be at the singular point where $R=0$. From Eq.~\eqref{eq:G_reduced} we see that the corresponding value of $g$ is $0$. 
In this case the topological circle is not smooth. Away from the singular point $R=0$, the points on this curve correspond to circular orbits of the action of $L_z$  giving together with the fixed point of the action $L_z$ at $R=0$ a pinched 2-torus where the pinch is a focus-focus singular point in the space reduced by the flow of $H$. Reconstructing the reduction by the flow of $H$ results in the  product of a pinched 2-torus and a circle in the original full phase space $T^*\mathbb{R}^3$.

The minimal value of $G$ attained at the singular point $R=E/\omega$ can be obtained from Eq.~\eqref{eq:G_reduced} and gives again \eqref{eq:min_g_general_lz}.  
The maximal value of $G$ can be computed from the condition that $\nabla G$ and $\nabla C$ are dependent on $C=0$, 
where $\nabla$ is with respect to the coordinates on the reduced space $(R, X, Y)$.  Similarly to the computation of the maximal value of $g$ for fixed $E$ and $l_z$ in  subsection~\ref{sec:bifdiag_gen} this leads to a cubic equation.  
The critical energy at which the focus-focus singular point comes into existence corresponds to the collision of the tangency that gives the maximal value of $g$ with the singular point $R=0$. As mentioned in subsection~\ref{sec:bifdiag_gen} this corresponds to a Hamiltonian Hopf bifurcation. The critical energy can be computed from comparing the slope of the upper branch of the slice $Y=0$ of $C=0$ at $R=0$ which is $2E$ with the 
slope of $G=0$ at $R=0$ which is $2a^2\omega^2-2E$. Equating the two gives the value $E=\frac12 \omega^2 a^2$ that we already found in 
subsection~\ref{sec:bifdiag_gen}.

%

\subsection{Symplectic volume of the reduced phase space}

\begin{figure}
\setlength{\unitlength}{1mm}
\begin{picture}(170,60)(0,0)
\put(5,0){\includegraphics[width=5cm]{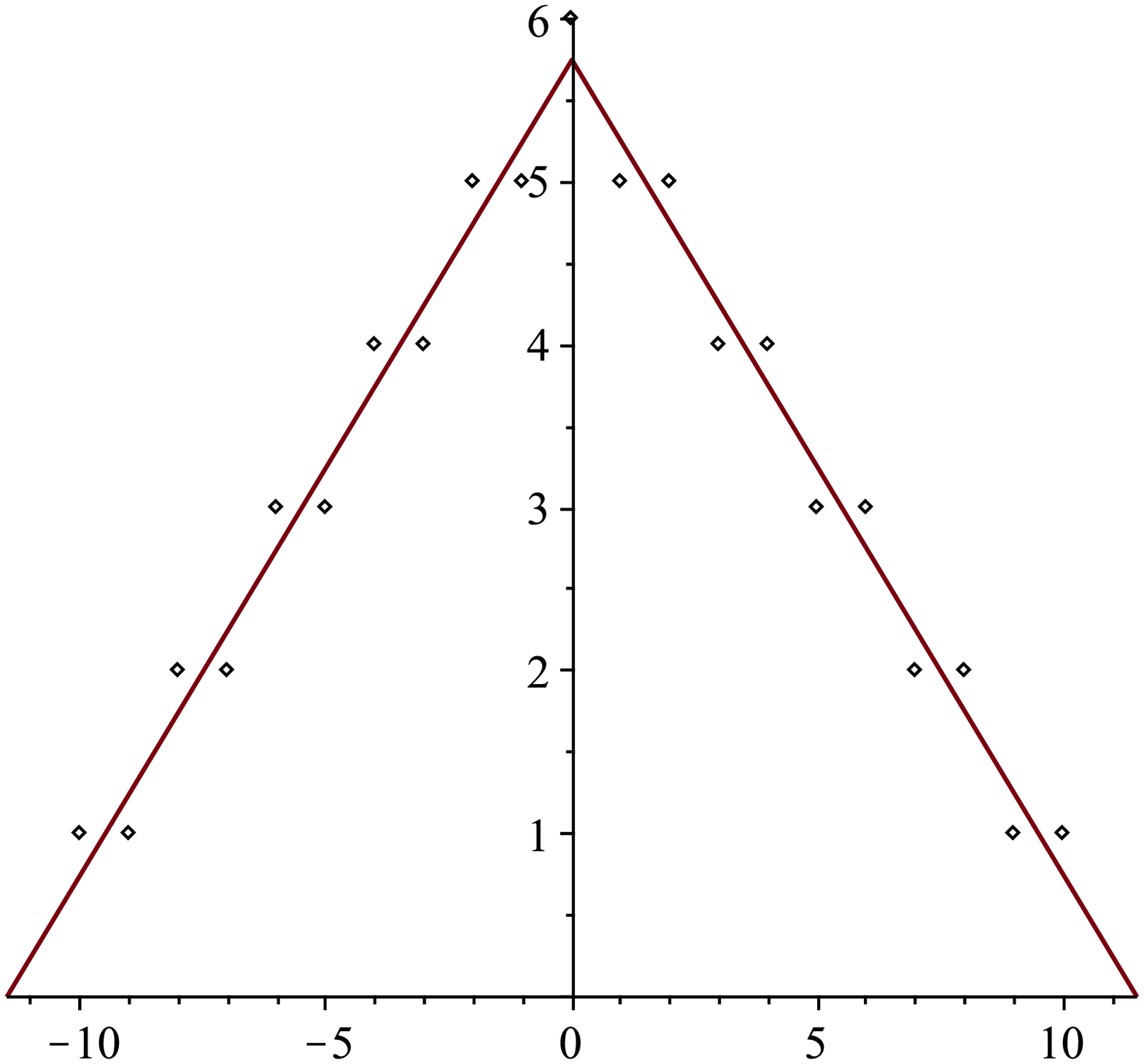}}
\put(0, 50){(a)}
\put(15,43){$\mathrm{Vol}_{E,l_z}$}
\put(27, 0){$l_z/\hbar$}
\put(60,0){\includegraphics[width=5cm]{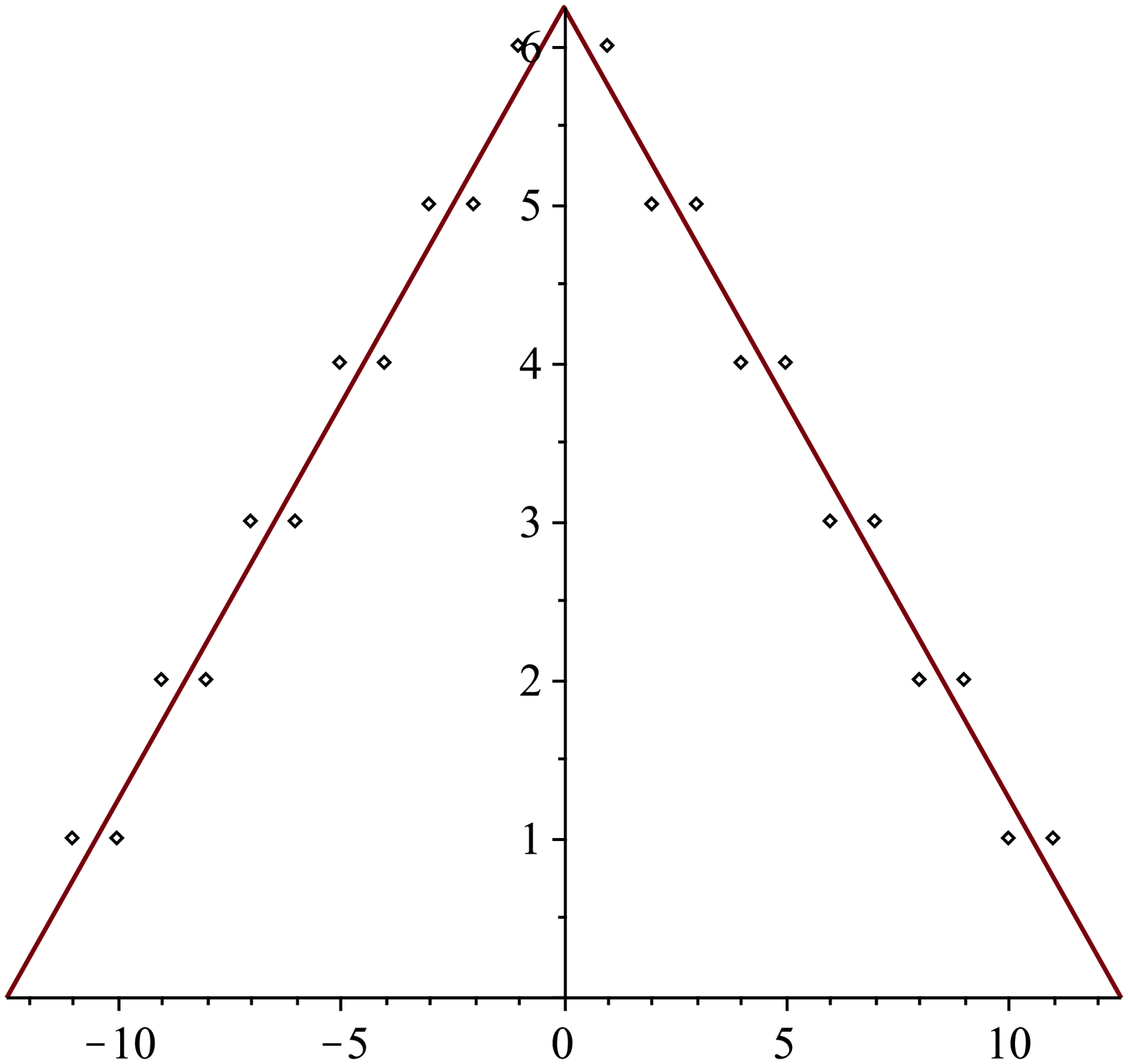}}
\put(55, 50){(b)}
\put(70,43){$\mathrm{Vol}_{E,l_z}$}
\put(82, 0){$l_z/\hbar$}
\put(117,0){\includegraphics[width=5cm]{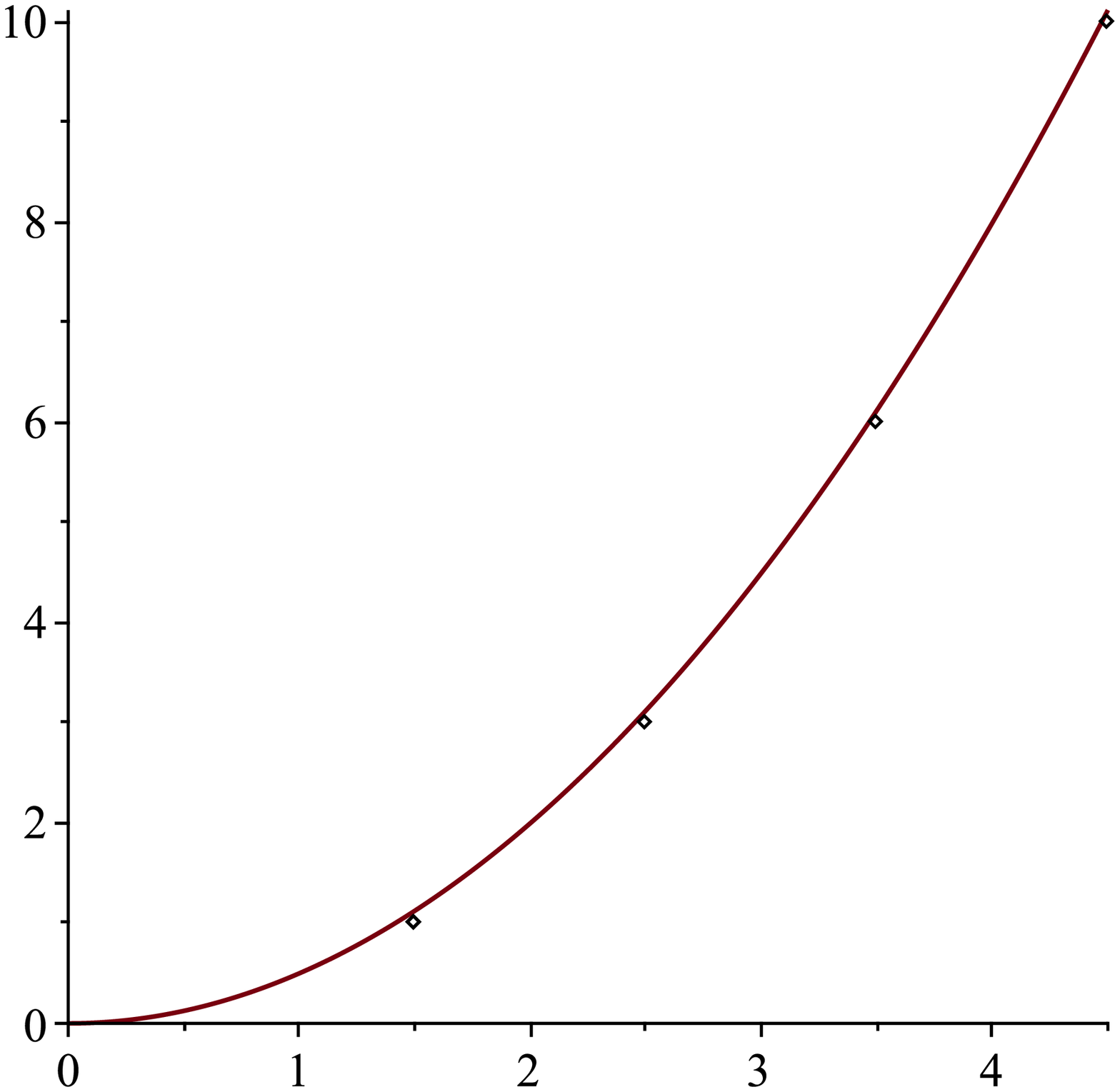}}
\put(110, 50){(c)}
\put(112,43){$N(E)$}
\put(155, 0){$E/(\hbar \omega)$}
\end{picture}
\caption{$\mathrm{Vol}_{E,l_z} $  versus $m=l_z/\hbar$  for $E=\hbar \omega (n+3/2)$ and corresponding exact number of states (dots) for the even integer $n=10$ (a) and the odd integer $n=11$ (b).
(c) The area $N(E)$ under graphs of the form in (a) and (b) divided by $2\pi\hbar^2$ versus $E$ and the corresponding exact number of states (dots) at energies of $\hbar\omega(n+3/2)$.
}\label{fig:WeylLaw}
\end{figure}

It follows from the Duistermaat-Heckman Theorem \cite{DuistermaatHeckman1982} that the symplectic volume (area) of the reduced phase space defined by $C = 0$ has a piecewise linear dependence on the global action $L_z$. 
Indeed, introducing cylinder coordinates to parametrize the reduced phase space $C = 0$ as
$X = f(R) \sin\theta$ and $Y=f(R) \cos\theta$ we see from 
$
  \{ \theta, R \} = 2,
$ 
that the symplectic form on $C=0$ is
$
  \frac12 d\theta \wedge dR.
$ 
 Integrating the symplectic form over the reduced space $C=0$ gives
the symplectic volume
\begin{align*}
  \mathrm{Vol}_{E,l_z} = \frac{\pi}{\omega}  (E- \omega |l_z|),
\end{align*}
for fixed $E \ge \omega |l_z|$. 
It follows from Weyl's law that $ \mathrm{Vol}_{E,l_z} /(2\pi \hbar) = (E-\omega |l_z|)/(2\hbar\omega)$ 
gives the mean number of quantum states for fixed $E$ and $l_z$ (see \cite{Ivrii2016} for a recent review). Indeed inserting $E=\hbar\omega (n+3/2)$ and $l_z=\hbar m$ we get 
$ \mathrm{Vol}_{E,l_z} /(2\pi \hbar) = (n+3/2 -|m|)/2$. Counting the exact number of states for fixed $n$ and $m$ 
which is most easily done using the separation with respect to spherical coordinates (see the Introduction)
we get $ (n+2-|m|)/2$ if $n-|m|$ is even and $ (n+1-|m| )/2$ if  $n-|m|$ is odd. 
We see that Weyl's law is interpolating between the even and the odd case, see Fig.~\ref{fig:WeylLaw}(a).

The area under the graph of $\mathrm{Vol}_{E,l_z}$ as a function of $l_z$ for fixed $E=\hbar\omega(n+3/2)$ is $\pi \hbar^2 (n+3/2)^2 $. Dividing by the product of $2\pi \hbar$ and $\hbar$ (which is the distance between two consecutive quantum angular momenta eigenvalues $l_z$) gives $(n+3/2)^2/2$ which for $n\to\infty$ asymptotically agrees with the exact number of states 
$(n+1)(n+2)/2$, see Fig.~\ref{fig:WeylLaw}(b). 


\subsection{The limiting cases $a\to 0$ and $a\to \infty$}
\label{sec:LimitingCases}

From Eq.~\eqref{eq:G_Cart} we see  that for $a\to 0$, the separation constant $G$ becomes  the squared  total angular momentum, $\mathbf{L}^2 = L_x^2 + L_y^2 + L_z^2 $.  In the limit $a \to 0$ we thus obtain the Liouville integrable system given by $(H, L_z, |\mathbf{L}|^2)$ which corresponds to separation in spherical coordinates.
Note that the $a \to \infty$ limit of prolate spheroidal coordinates corresponds to parabolic coordinates, where the harmonic oscillator is not separable.
However, the scaled separation constant
\begin{equation}\label{eq:G_Cart_mod} 
\tilde{G}  = -\frac{1}{a^2} G = 2  (A_x + A_y) - \frac{1}{a^2} (L_x^2 + L_y^2 + L_z^2 ),
\end{equation}
has the well defined limit $2  (A_x + A_y) $ as $a\to \infty$. 
The limit $a \to \infty$ then leads to the Liouville integrable system $(H, L_z, 2(A_x + A_y))$.
The standard separation in Cartesian coordinates leads to the integrable system $(H, A_x, A_y)$.


The reduction by the flow of $H$ gives as the reduced space the compact symplectic manifold $\mathbb{C}P^2$, see Sec.~\ref{sec:separation}.
Then the map $(A_x, A_y)$ associated with separation in Cartesian coordinates defines an effective toric action on $\mathbb{C}P^2$. 
The image of  $\mathbb{C}P^2$ under $(A_x, A_y)$ is therefore a Delzant polygon which is a convex polygon with special properties \cite{Delzant1988}, see Fig.~\ref{fig:polygon3}(a).

Similarly the map $(L_z,\frac{1}{\omega}(A_x + A_y)):\mathbb{C}P^2\to \mathbb{R}^2$ associated with separation in prolate spheroidal coordinates in the limit  $a\to \infty$ also defines a toric, non-effective, action and its image is the convex, non-Delzant, polygon shown in Fig.~\ref{fig:polygon3}b. We here have scaled the separation constant in such a way that the $S^1$ actions associated with the flows of $L_z$ and $\frac{1}{\omega}(A_x + A_y)$ have the same period.

The image of the map $(L_z, |\mathbf{L}|):\mathbb{C}P^2\to  \mathbb{R}^2$ associated with the limit $a\to 0$ and separation in spherical coordinates also gives the same polygon as in the previous case, see Fig.~\ref{fig:polygon3}c. However, whereas here $L_z$ is a global $S^1$ action this is not the case for $|\mathbf{L}|$ whose Hamiltonian vector field is singular at points with $\mathbf{L}=0$. 
Because of this singularity $(L_z, |\mathbf{L}|)$ is not the moment map of a global toric action. Whereas the image is a convex polygon the singularity manifests itself when considering the joint quantum spectrum of the operators associated with the classical constants of motion. Whereas these form rectangular lattices in Figs.~\ref{fig:polygon3}(a) and (b) with lattice constants $\hbar$ this is not the case in Fig.~\ref{fig:polygon3}(c) where the distance between consecutive lattice layers is not constant in the vertical direction.


\begin{figure}
\setlength{\unitlength}{1mm}
\begin{picture}(170,50)(0,0)
\put(5,11){\includegraphics[width=3.3cm]{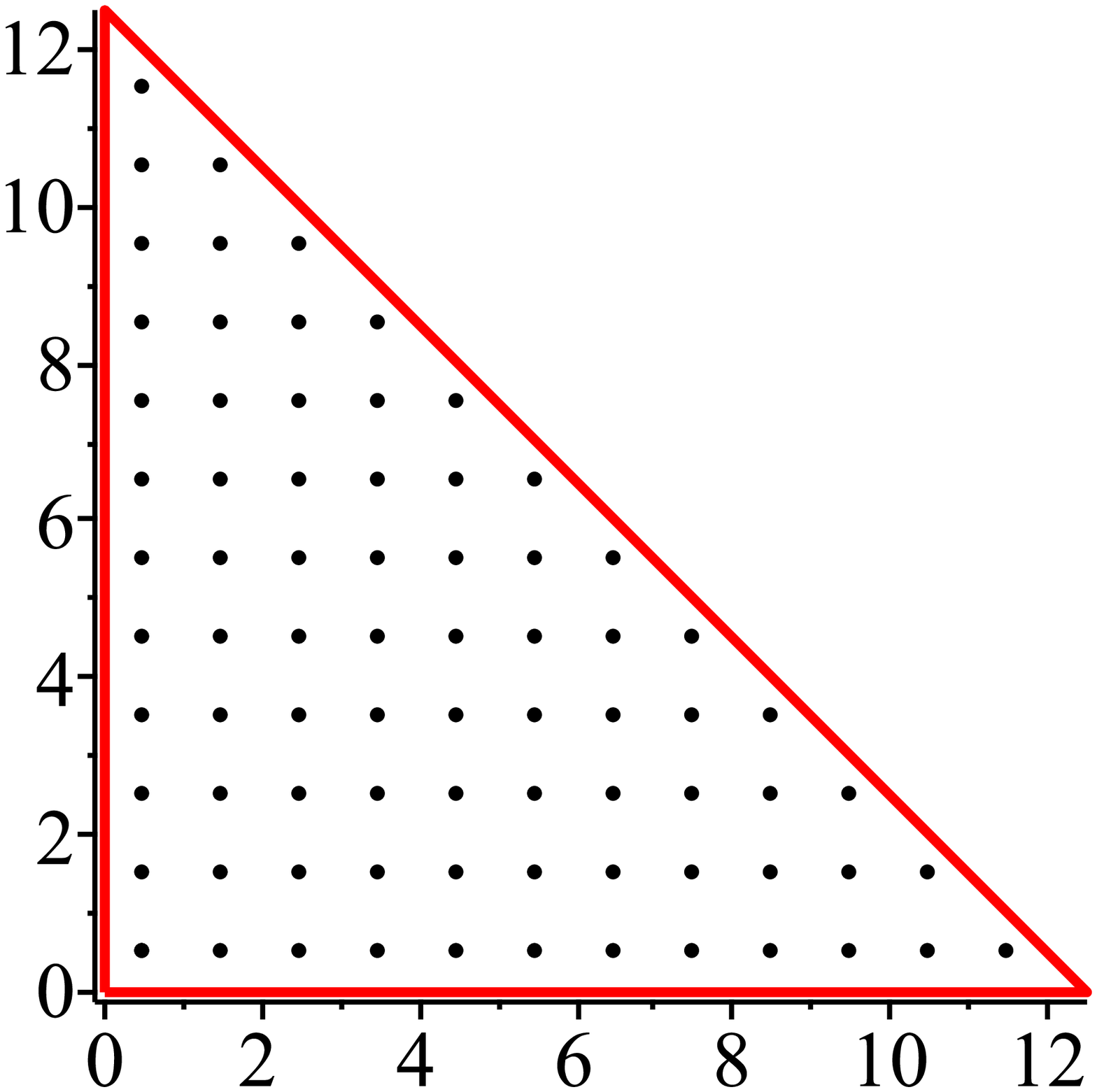}}
\put(0, 50){(a)}
\put(2,40){$a_y$}
\put(33, 10){$a_x$}
\put(45,0){\includegraphics[width=6cm]{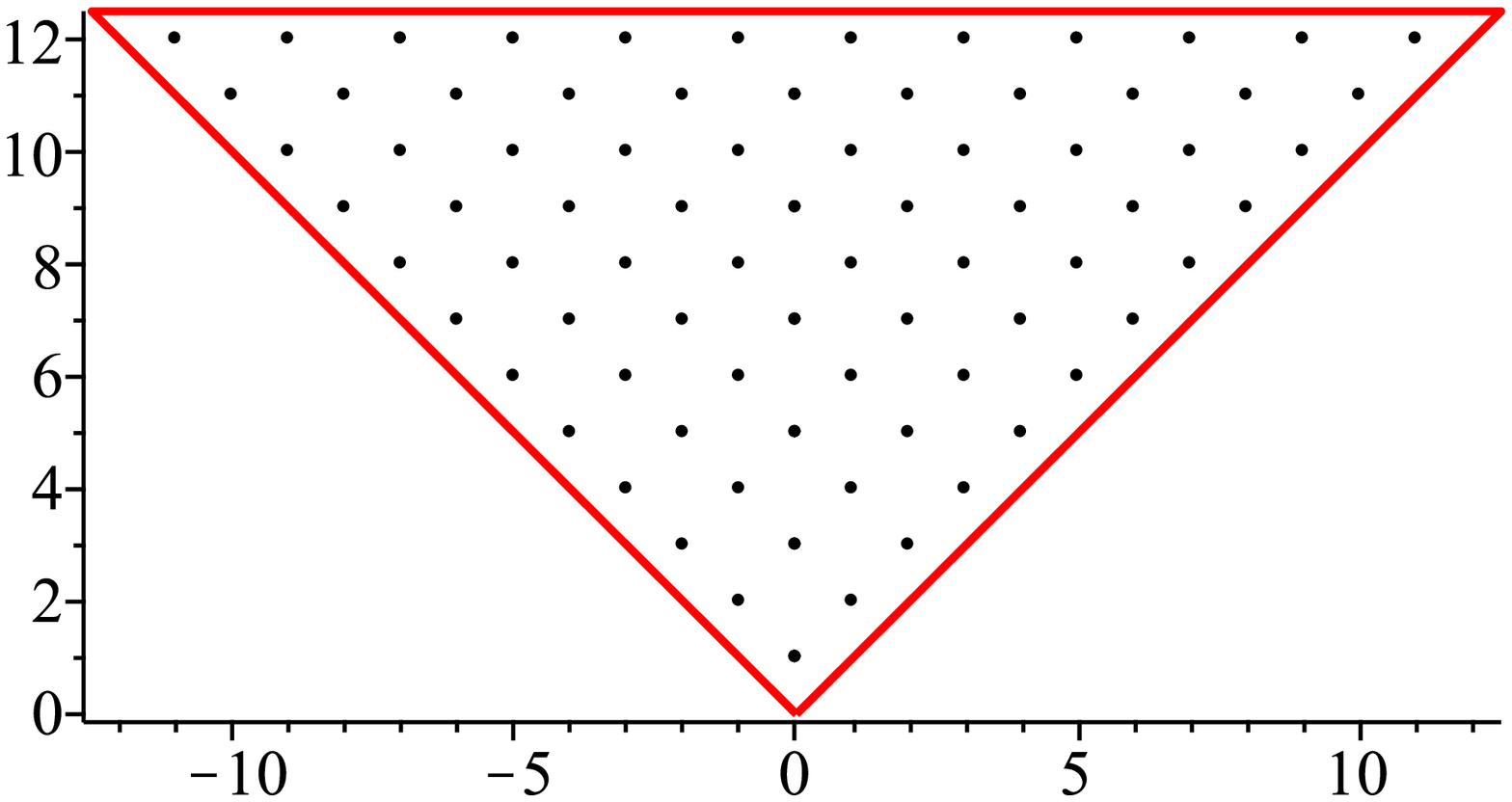}}
\put(39, 50){(b)}
\put(43,22){\rotatebox{90}{$(a_x+a_y)/\omega$}}
\put(77, 10){$l_z$}
\put(111,0){\includegraphics[width=6cm]{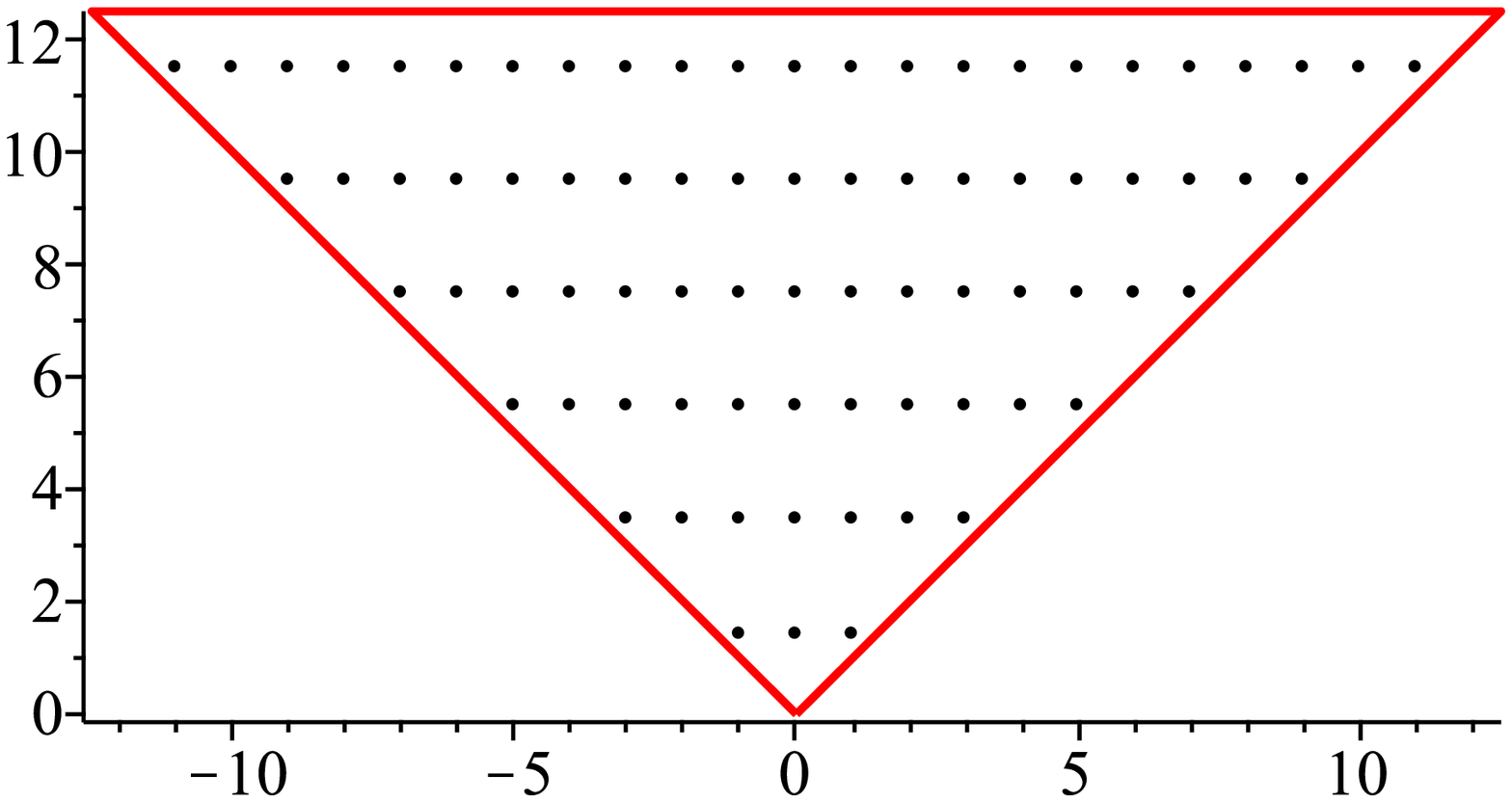}}
\put(108, 50){(c)}
\put(111,40){$l $}
\put(143, 10){$l_z$}
\end{picture}
\caption{The images of  different maps of integrals $\mathbb{C}P^2\to \mathbb{R}$ where $\mathbb{C}P^2$ is the energy level set of the 
harmonic oscillator reduced  by the flow of the Hamiltonian $H$. 
(a) The map of integrals $(A_x, A_y)$ associated with separation in Cartesian coordinates, where we denote the values of the functions $A_k$ by $a_k$, $k=x,y$.The image is enclosed by the triangle with corners $(0,0)$, $(0,E)$ and $(E,0)$.
(b) The map of integrals  $(L_z, \frac{1}{\omega}(A_x+A_y))$ associated with the limit $(a \to \infty)$ when separating in prolate spheroidal coordinates.  The image is  enclosed by the triangle with corners $(0,0)$, $(E/\omega,E/\omega)$ and $(-E/\omega,E/\omega)$.
(c) The map of integrals $(L_z, |\mathbf{L}| )$ associated with separation in spherical coordinates and the limit $(a\to 0)$ in prolate spheroidal coordinates. Here $l$ denotes the value of the function $|\mathbf{L}| $. The image is  enclosed by the triangle with corners $(0,0)$, $(E/\omega,E/\omega)$ and $(-E/\omega,E/\omega)$. The dots mark the joint spectrum of the corresponding quantum operators. The energy is chosen to be $E=\omega\hbar(n+3/2)$ with $n=11$. The values of $\hbar$ and $\omega$ are chosen as 1.
} \label{fig:polygon3}
\end{figure}

\rem{
\begin{figure}
\raisebox{5cm}{(a)} \raisebox{1cm}{\includegraphics[width=3.5cm]{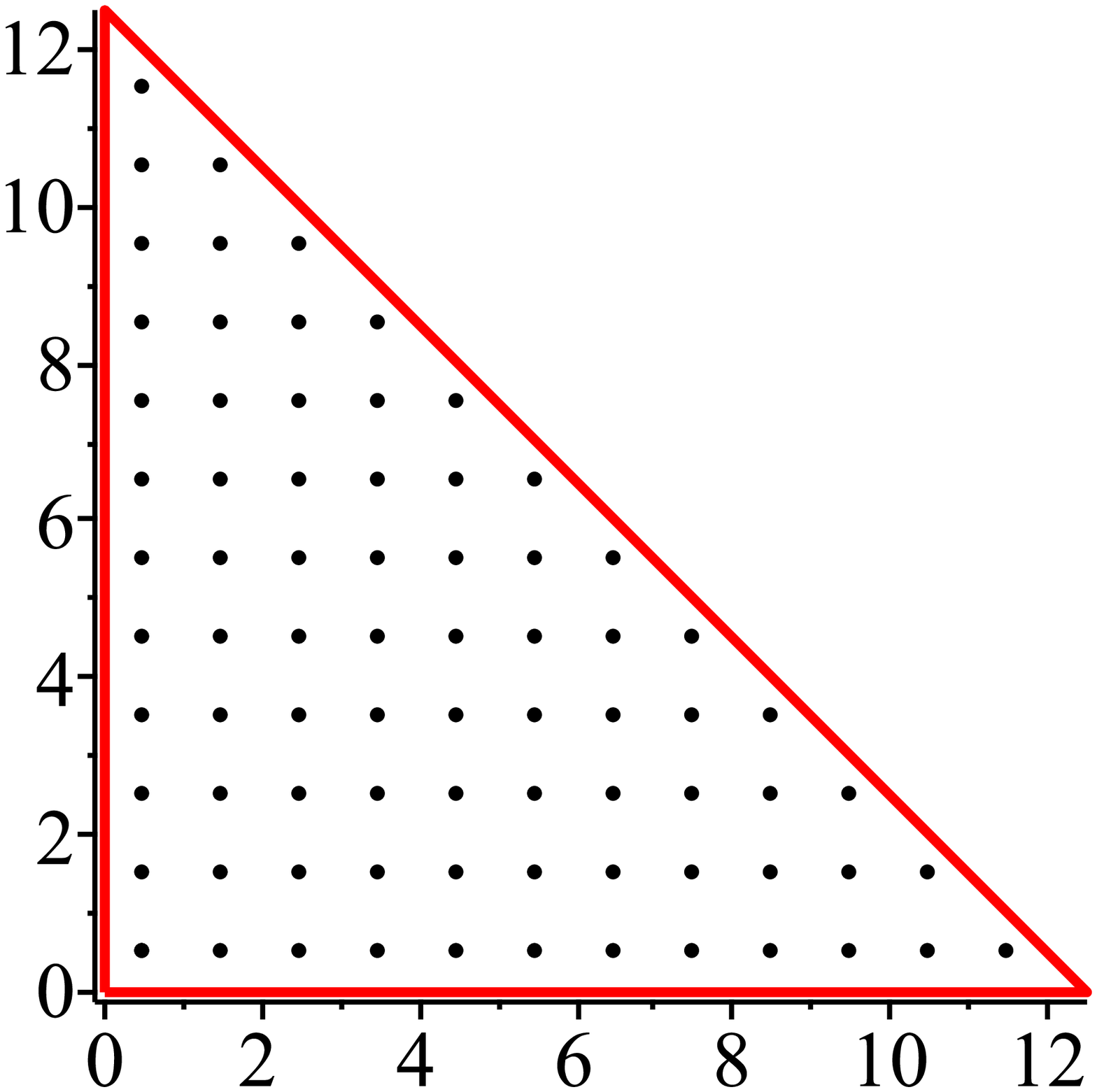}}
\raisebox{5cm}{(b)} \includegraphics[width=6cm]{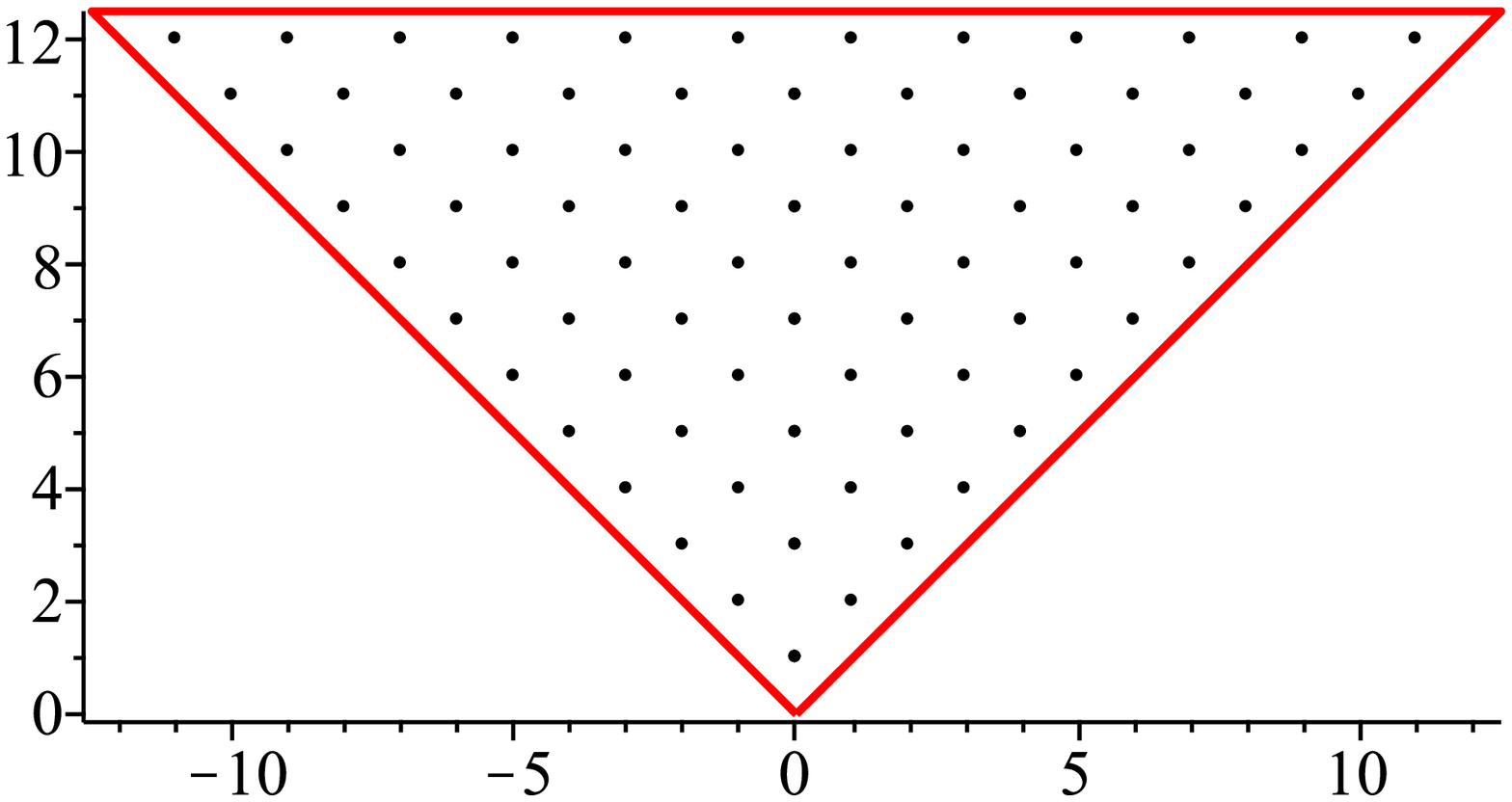}
\raisebox{5cm}{(c)} \includegraphics[width=6cm]{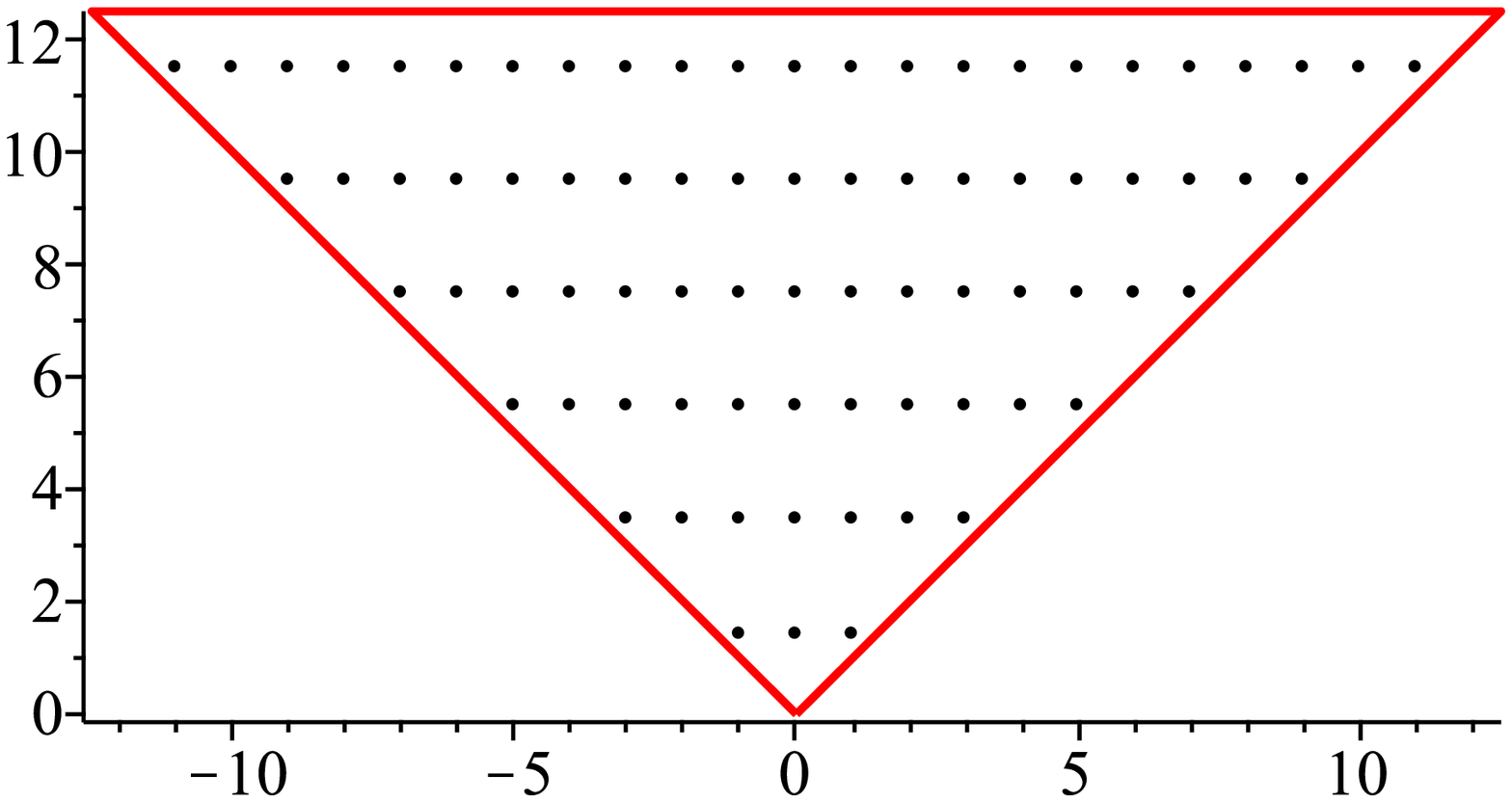}
\caption{The images of  different maps of integrals $\mathbb{C}P^2\to \mathbb{R}$ where $\mathbb{C}P^2$ is the phase space of the 
harmonic oscillator reduced  by the flow of the Hamiltonian $H$. 
(a) The map of integrals $(A_x, A_y)$ associated with separation in Cartesian coordinates. The image is enclosed by the triangle with corners $(0,0)$, $(0,E)$ and $(E,0)$.
(b) The map of integrals  $(L_z, \frac{1}{\omega}(A_x+A_y))$ associated with the limit $(a \to \infty)$ when separating in prolate spheroidal coordinates.  The image is  enclosed by the triangle with corners $(0,0)$, $(E/\omega,E/\omega)$ and $(-E/\omega,E/\omega)$.
(c) The map of integrals $(L_z, |\mathbf{L}| )$ associated with separation in spherical coordinates and the limit $(a\to 0)$ in prolate spheroidal coordinates. The image is  enclosed by the triangle with corners $(0,0)$, $(E/\omega,E/\omega)$ and $(-E/\omega,E/\omega)$. The dots mark the joint spectrum of the corresponding quantum operators. The energy is chosen to be $E=\omega\hbar(n+3/2)$ with $n=11$. The values of $\hbar$ and $\omega$ are chosen as 1.
} \label{fig:polygon3}
\end{figure}
} 


\section{Quantum monodromy}
\label{sec:sepQM}

In this section we discuss the implications of the monodromy discussed in the previous section on the joint spectrum of the quantum mechanical version of the isotropic oscillator which is described by the operator
\[
       \hat{H} = -\frac{\hbar^2}{2} \nabla^2 + \frac{\omega^2}{2} (x^2 + y^2 + z^2) \,.
\]

In prolate spherical coordinates the Schr\"odinger equation becomes
\[
-\frac{\hbar^2}{2} \left\{ 
\frac{1}{a^2(\xi^2-\eta^2)}
\left[ 
\frac{\partial}{\partial \xi} \left((\xi^2-1) \frac{\partial \Psi}{\partial \xi} \right) 
+ \frac{\partial}{\partial \eta} \left((1-\eta^2) \frac{\partial \Psi}{\partial \eta} \right) 
\right]
+ \frac{1}{a^2(\xi^2-1)(1-\eta^2)} \frac{\partial^2 \Psi}{\partial \phi^2}
\right\}
+ \frac{\omega^2}{2} a^2 (\xi^2 + \eta^2 -1 ) \Psi = E \Psi.
\]
Separating the Schr\"odinger equation in prolate spheroidal coordinates works similarly to the classical case discussed in Sec.~\ref{sec:separation}. 
The separated equations for $\eta$ and $\xi$ are
\begin{equation} \label{eq:Ppsi}
  - \hbar^2 \frac{1}{1 - s^2} \frac{d}{ds} \left[(1 - s^2) \frac{d\psi}{ds} \right] = \frac{P(s)}{(1-s^2)^2} \psi \,,
\end{equation}
where $P(s)$ is again the polynomial that we defined for the classical case in Eq.~\eqref{eq:def_P(s)}, 
with $l_z = \hbar m$.
This is the spheroidal wave equation with an additional term proportional to $\omega^2$ coming from the potential.
For $|s| < 1$ it describes the angular coordinate $\eta$, and for $s > 1$ the radial coordinate $\xi$ 
of spheroidal coordinates.

Analogously to the classical case the separation constant $g$ corresponds to the eigenvalue of the operator 
\begin{equation}\label{eq:G_Cart_quantum} 
\hat G  =  {\hat L}_x^2 + {\hat L}_y^2 + {\hat L}_z^2  - 2 a^2 ({\hat A}_x + {\hat A}_y)\, ,
\end{equation}
where for $k=x,y,z$,   the ${\hat L}_k$,  are the components of the standard angular momentum operator, and the ${\hat A}_{k}=-\frac12 \hbar^2 \partial^2_k + \frac12 \omega^2 k^2$  are the Hamilton operators of one-dimensional harmonic oscillators.


A WKB ansatz shows that the joint spectrum of the quantum integrable system $(\hat H, {\hat L}_z,\hat G)$ associated with the separation in prolate spheroidal coordinates can be computed semi-classically from a Bohr-Sommerfeld quantization of the actions according to
$I_\phi =  \frac{1}{2\pi}  \oint p_\phi \,\ud \phi = \hbar m $,  
$I_\eta = \frac{1}{2\pi}  \oint p_\eta \,\ud \eta  = \hbar (n_\eta+\frac12) $ and $I_\xi =  \frac{1}{2\pi}   \oint p_\xi \,\ud \xi = \hbar(n_\xi + \frac12 )$ with $m \in \Z$
and non-negative quantum numbers $n_\eta$ and $n_\xi$. 
Using the calculus of residues it is straightforward to show that $E = I_\eta + I_\xi + | I_\phi |$.
Taking the derivative with respect to $l_z$ using $I_\phi=l_z$ shows that  
the actions $I_\eta$ and $I_\xi$ are not globally smooth  functions of the constants of motion $(E,g,l_z)$.
This is an indication that the quantum numbers do not lead to a globally smooth labeling of the joint spectrum.  
We will see this in more detail below.


\subsection{Confluent Heun equation}

It is well known that the spheroidal wave equation can be transformed into the confluent Heun equation
\cite{NIST:DLMF}. Adding the harmonic potential adds additional terms that dominate at infinity, and 
so a different transformation needs to be used to transform \eqref{eq:Ppsi} into the Heun equation. 
We change the independent variable $s$ in \eqref{eq:Ppsi} to $u$ by $s^2 = u$ and the dependent variable to $y$ by 
$y(u) = \exp( a^2 \omega u/(2 \hbar) ) ( 1 - u)^{-m/2} \psi(s)$ which leads to 
\[
    y'' + \left( - \frac{a^2\omega}{\hbar} - \frac{m+1}{1-u} + \frac{1}{2 u}\right) y'  + Q y = 0\,,
\] 
where
\[  
    Q = \frac{ g -  \hbar^2 m(m+1)}{4 \hbar^2 u ( 1 - u)} + 
       \frac{ a^2}{2\hbar^2} \left(  \frac{\hbar (m+1)}{ 1-u} + \frac{E-\tfrac12 \hbar \omega}{  u} \right)  \,.
 \]
This is a particular case of the confluent Heun equation, with regular singular 
points at 0 and 1, and an irregular singular point of rank 1 at infinity.
Each regular singular point has one root of the indicial equation equal to zero, 
so we may look for a solution of the form $y(u) = \sum_k a_k u^k = \sum_k b_{2k} s^{2k}$. 
This leads to the three-term recursion relation for the coefficients
\[
    b_{k-2} A_{k-2} + b_k B_k - b_{k+2}  \hbar^2 (k+1)(k+2)  = b_k ( g + 2 a^2 E)\,,
\]
where $k$ is an even integer and
\begin{align*}
A_{k-2} & = 2 a^2 (E - \hbar \omega (m + k -  \tfrac12 )), \\  
B_k & =  a^2 \hbar \omega (2k+1) + \hbar^2 (m+k)(m+k+1) \,.
\end{align*}
If we require that $y(u)$ is polynomial of degree $d$,
we need to require that for $k = 2 d + 2$ the coefficient $A_{k-2}$ vanishes, 
and hence the quantisation condition
\[
    E = \hbar \omega \left(  m + 2 d + \tfrac 32\right) ,
\]
with principal quantum number $n = m + 2 d$ is found.
Fixing $E$ to some half-integer the spectrum of the tridiagonal matrix $M$ obtained
from the three-term recurrence relation
determines the spectrum of $ g + 2 a^2 E$.
In the limit $a \to 0$ the spectrum 
becomes $n(n+1), ..., m(m+1)$. Note that fixing the energy and 
allowing all possible degrees $d$ makes $m$ change in steps of 2.
Since $m$ in fact changes in steps of 1 there must be additional solutions.

The regular singular point at $u=0$ has another regular solution 
with leading power $\sqrt{u} = s$, so that we make the Ansatz $y(u) = \sqrt{u} \sum_0 a_k u^k = \sum_0 b_{2k+1} s^{2k+1}$, 
which leads to an odd function in $s$.
The same three-term recursion relation holds as above, except that now the index $k$ is odd.
For $a\to 0$ the spectrum is $n(n+1), \dots, (m+1)(m+2)$, as before in steps of 2 in $m$.

We note that in the spherical limit $a\to 0$ the Heun equation reduces to the associated Laguerre equation
with polynomial solutions $L_{(n-l)/2}^{l+1/2}(s^2)$ when $g = l(l+1)$.


\subsection{Algebraic computation of the joint spectrum}

Instead of starting from the spheroidal wave equation wave equation as illustrated in the previous subsection one can directly compute the joint spectrum algebraically by using creation and annihilation operators. As we will see this gives explicit expressions for the entries of a tri-diagonal matrix whose eigenvalues give the spectrum of $\hat G$ for fixed $E$ and $l$. 

Instead of the usual creation and annihilation operators of the
harmonic oscillator we use operators that are written in the set of coordinates $(z_1, z_2, z_3)$ introduced in Sec.~\ref{sec:reduction}. 
The transformation to the new coordinates diagonalises $\hat L_z$ and at the same time keeps $\hat H$ diagonal, so that 
\[
   \hat  H = \hbar \omega ( a_1^\dagger a_1 + a_2^\dagger a_2 + a_3^\dagger a_3 + \tfrac32), \quad
   \hat L_z = \hbar (  a_1^\dagger a_1 - a_2^\dagger a_2 ) \,.
\]
and the operator $\hat R$ corresponding to the classical  $R$ in Eq.~\eqref{eq:def_R} reads
\[
   \hat R = \hbar ( a_1^\dagger a_1 + a_2^\dagger a_2 + 1) \,.
\]
The operator $\hat X$ corresponding to $X$ in Eq.~\eqref{eq:def_X} is of higher degree, and thus care needs to be taken with the order of operators.
The classical $X$ can be written as $X = \frac{1}{2} \omega (  z_1 z_2 \bar z_3^2 + \bar z_1 \bar z_2 z_3^2)$.
We also need to preserve the relation (for operators!)
$\omega \hat L^2 = \omega \hat L_z^2 + 2 ( \hat H - \omega \hat R) \hat R + \hat X$, cf.~Eq.~\eqref{eq:def_X},
and this leads to
\[
    \hat X = 2 \hbar^2 \omega \left( a_1^\dagger a_2^\dagger a_3^2 + a_1 a_2 ( a_3^\dagger)^2 - \frac12 \right) \,.
\]
With these expressions matrix elements can be computed. Denote a state
with three quantum numbers associated to the creation and annihilation operators $a_i$ and $a_i^\dagger$, $i=1,2,3$, 
by $\ket{k_1, k_2, k_3}$, such that 
\begin{align*}
   a_1^\dagger \ket{k_1, k_2, k_3} & = \sqrt{ n_1 + 1}  \ket{k_1+1, k_2, k_3}, \\
   a_1  \ket{k_1, k_2, k_3}              & = \sqrt{ n_1 }  \ket{k_1-1, k_2, k_3}, \text{ for } k_1 \ge 1, \\
   a_1    \ket{0, k_2, k_3}    & = 0
\end{align*}
and similar relations for $a_2$ and $a_3$. This allows to verify
\begin{align*}
    \hat H \ket{k_1, k_2, k_3}      &= \hbar \omega (k_1 + k_2 + k_3 + \tfrac32) \ket{k_1, k_2, k_3}\,, \\
    \hat L_z \ket{k_1, k_2, k_3}   &= \hbar (k_ 1 - k_2) \ket{k_1, k_2, k_3} \,.
\end{align*}
In terms of the quantum numbers $(k_1,k_2,k_3)$ the principal and magnetic quantum numbers are  $n = k_1 + k_2 +k_3$ and 
$m = k_1 - k_2$, respectively. The space of states with fixed $n$ and fixed
$m$ is the span of the states of the form
\[
   \ket{k} := \ket{k, k - m, n + m - 2k } , \quad \max(0, m) \le k \le \frac12 ( n + m) \,.
\]
Now the non-zero matrix elements of
\begin{align*}
  \hat G 
  = \hat{\mathbf L}^2 - 2 a^2 \omega \hat R
  = \hat{L}_z^2 - 2 \hat R^2 - \frac{2}{\omega} (a^2 \omega^2 - \hat H) \hat R + \frac{1}{\omega} \hat X,
\end{align*}
are given by 
\begin{align*}
    \bra{k} \hat G \ket{k} & = 2\hbar a^2 \omega ( m - 1 - 2k) - \hbar^2( 2+ 8 k(1+k)   -4m -8k m + m^2  )  + \frac{\hbar^2}{\omega} (1+2k-m)(3+2n)\\
    \bra{k} \hat G \ket{k+1} &= 2 \hbar^2 \sqrt{ (k+1)(k+1-m)(n-1+m-2k)(n+m-2k)}.
\end{align*}

\begin{figure}
 \includegraphics[width=0.7\linewidth]{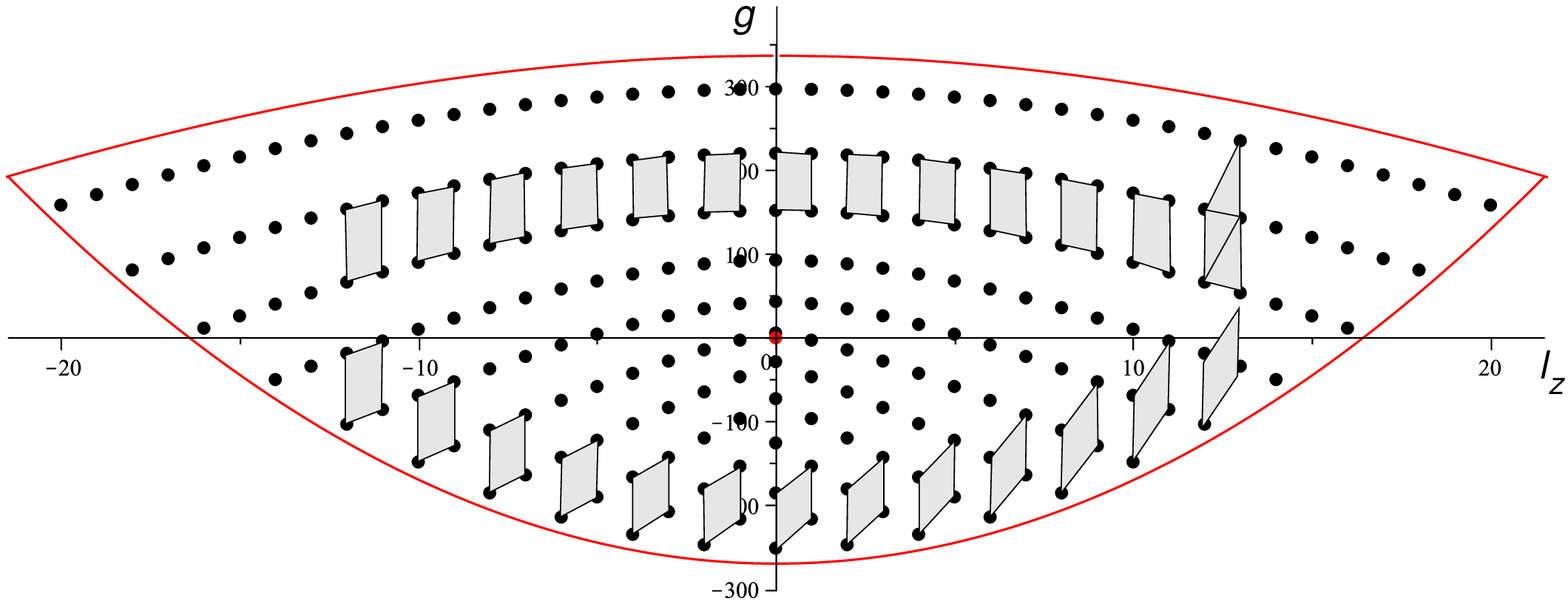}
\caption{Joint spectrum $(l_z,g)$ of $(\hat L_z, \hat G)$ (black dots) and 
classical critical values (red), 
for $n=20$, $\omega=1$, $\hbar=1$, and $a = 3/2$. There are $(n+1)(n+2)/2$ joint states.
The joint spectrum locally has a lattice structure which globally has a defect as can be seen from transporting a lattice cell around the isolated critical value at the origin.
} \label{fig:MoMap_mono}
\end{figure}

The resulting joing spectrum of $({\hat L}_z,\hat G )$ for a fixed $n$ is shown in Fig.~\ref{fig:MoMap_mono} for a choice of parameters such that the energy $E$ is above the threshold value $\frac12 \omega^2 a^2$ for the occurrence of monodromy. As to be expected from the Bohr-Sommerfeld quantization of actions the spectrum locally has the structure of a regular grid.
Globally however the lattice has a defect as can be seen from transporting a lattice cell along a loop that encircles the isolated critical value of the energy momentum map at the origin.

\begin{figure}
\raisebox{3cm}{(a)} \includegraphics[width=0.45\linewidth]{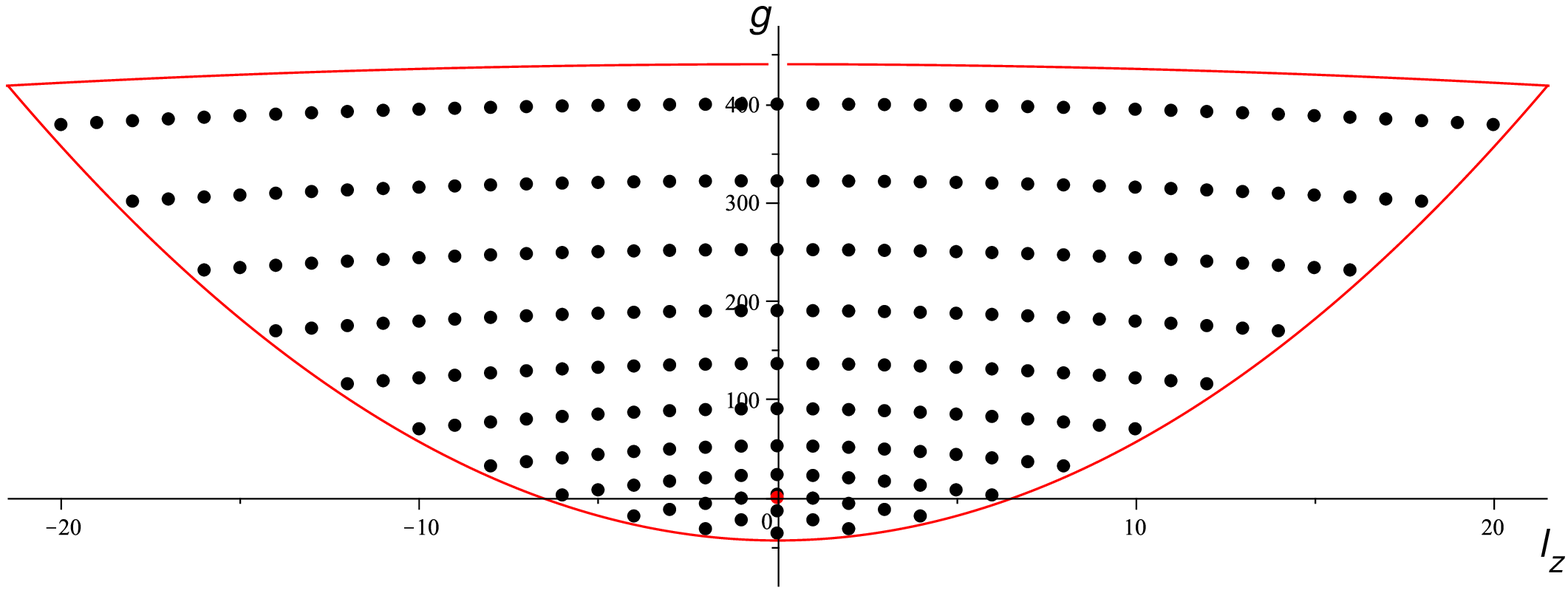}
\raisebox{3cm}{(b)} \includegraphics[width=0.45\linewidth]{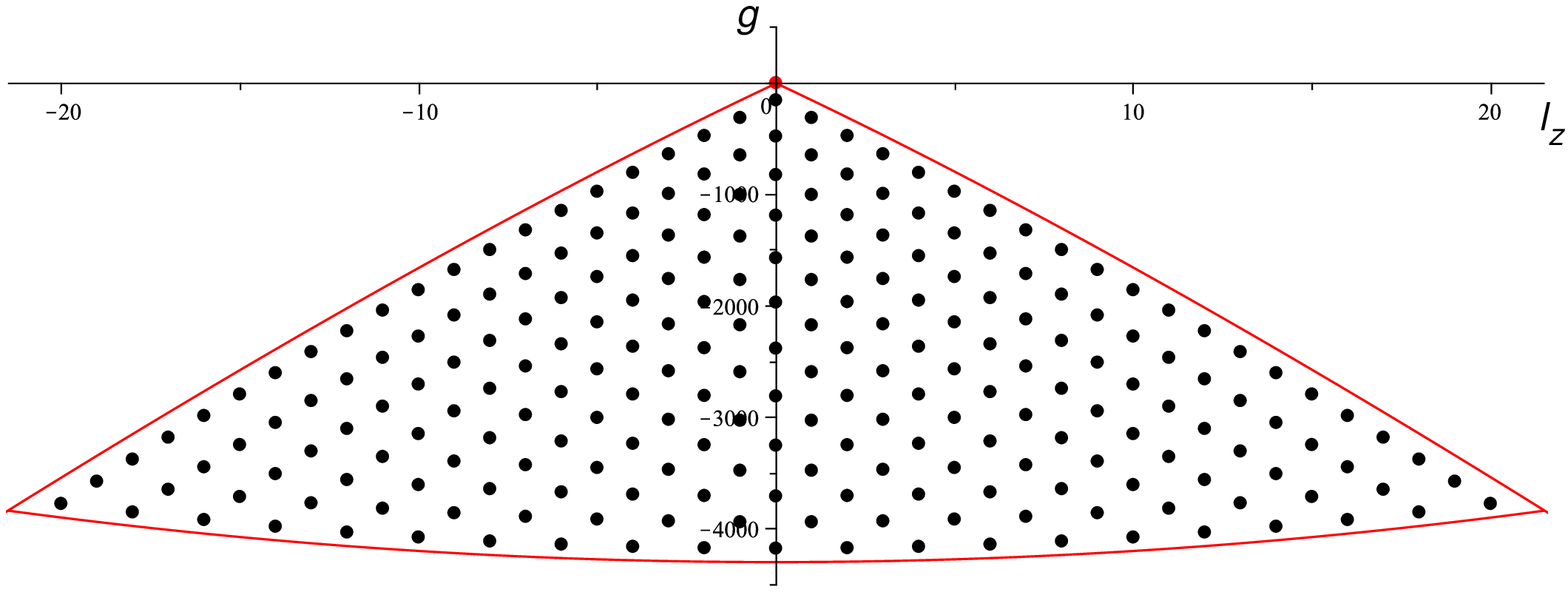}
\caption{Joint spectrum $(l_z,g)$ of $(\hat L_z, \hat G)$ (black dots) and 
classical critical values (red) for 
$a = 1$ (a) and $a=10$ (b)  and otherwise same parameters as in  Fig.~\ref{fig:MoMap_mono}.
} \label{fig:MoMap}
\end{figure}

In Fig.~\ref{fig:MoMap} the joint spectrum  of $({\hat L}_z,\hat G)$ is shown for fixed $n$ and  a small and large value of $a$, respectively. 
As discussed in Sec.~\ref{sec:LimitingCases}, in the limits $a\to 0$ and $a\to\infty$ (and in the latter case changing to $\tilde{G}=-\frac{1}{a^2}G$) the images  become the polygones  shown in Fig.~\ref{fig:polygon3}.

\section{Discussion}
\label{sec:discussion}


It is interesting to compare the two most important super-integrable systems, the Kepler problem and the harmonic oscillator, in the light of our analysis.
The Kepler problem has symmetry group $\mathrm{SO}(4)$ and reduction by the Hamiltonian flow leads to a system on $S^2 \times S^2$ \cite{Moser1970}.
The 3-dimensional harmonic oscillator has symmetry group $SU(3)$ 
and reduction by the Hamiltonian flow leads to a system on $\mathbb{C}P^2$. 

Separation of both systems, the Kepler problem and the harmonic oscillator in 3 dimensions, in prolate spheroidal coordinates leads to Liouville integrable systems that are of toric type for sufficiently large $a$. 
Here the technical meaning of toric type is that they are integrable systems with a global $T^n$ action for $n$ degrees of freedom, which implies that all singularities are of elliptic type. 
To a toric system is associated the image of the momentum map of the $T^n$ action, and this is a Delzant polytope, a convex polytope with special properties \cite{Delzant1988}.
The Delzant polytope for the $T^2$ action of the reduced Kepler system on $S^2 \times S^2$ is a square (take the limit $a\to\infty$ in
Fig.~4 in \cite{DullinWaalkens2018})
while the Delzant polytope for the $T^2$ action of the reduced harmonic oscillator on $\mathbb{C}P^2$ is an isosceles right triangle, see Fig.~\ref{fig:polygon3}(a).
It is remarkable that the two simplest such polytopes appear as reductions from the Kepler problem and 
from the harmonic oscillator. We note, however, that the harmonic oscillator  as opposed to the Kepler problem does not separate in parabolic coordinates.
This is related to the fact that for the separation of the Kepler problem in prolate spheroidal coordinates, the origin is in a focus point, while for the oscillator the origin is the midpoint between the foci. 

For decreasing family parameter $a$, both systems become semi-toric \cite{Pelayo2009,EFSTATHIOU2017104} through a supercritical Hamiltonian Hopf bifurcation.
It thus appears that the reduction of super-integrable systems by the flow of $H$ leads to natural and important examples of toric and semi-toric systems on compact symplectic manifolds.

\rem{
\section{Remarks / ToDo}

\begin{itemize}
\item What is the limit of $I_\xi$ and $I_\eta$ when $a \to 0$?
\item \RESOLVED{Introduce canonical variables on reduced phase space defined by Casimirs and compute the volume
of the reduced phase space, plot it as a function of $l$ (Duistermaat-Heckman). This will give a 
piecewise linear function, and should account for Weyl's law. For the HO the slope must be
less, since the number of states is less. KE: This is already done.}
\item Mention the Sturm-Liouville weights so that we know what is orthogonal, 
and check  what they are for spheroidal coordinates, probably $=1$ in the algebraic form?
\item check details of $a\to 0$ limit of Heun equation; what about $a\to \infty$?
\end{itemize}

\hrule

\begin{itemize}
\item insert comments and references to work on superintegrable systems in the introduction (see the hydrogen paper)
\item \RESOLVED{HW: it would be nice to explicitly see somewhere in the classical part that the reduced phase space obtained from reducing the flow of $H$ is indeed $CP^2$. KE: It is a non-trivial exercise to start with the invariants and their syzygies (which, by the way, we don't include in the paper) and show that the space they describe is $CP^2$. The standard proof is indirect: invariants and the syzygies are known to give the orbit space of the $H$-induced $S^1$ action, and $CP^2$ is \emph{defined} to be the orbit space of exactly this $S^1$ action, so the result follows. I have added some text in Sec. II concerning this.}
\item \RESOLVED{HW: note that in the classical part there is a separating constant $G$ and $\tilde{G}= - G/a^2$ (note the sign). It seems that $\tilde{G}$ is used in the quantum part. But still there is a sign issue in the separated Schrodinger equation. In the classical part all signs and factors like $a$ and $\omega$ are checked very carefully (here only in the systematic way of deriving the invariants this still needs to be done; see comment in the text). KE: I believe that signs are OK now}\ISSUE{except perhaps for the Heun equation}
\item \ISSUE{a quantum monodromy picture where a lattice cell is transported around the singularity is still missing. KE: For HW?}
\item it would be nice to have a plot of the wave functions and caustics like in the hydrogen paper...
\item should the actions be discussed in more detail? At the moment their non smoothness is only mentioned in the WKB approximation in the quantum part. Should we compute the monodromy matrix from smoothing the actions like in Irina's thesis?

\end{itemize}

} 


%

\end{document}